\documentclass[twocolumn,pra,superscriptaddress,aps]{revtex4-2}

\usepackage{graphicx,epstopdf}
\usepackage{dcolumn}
\usepackage{bm}

\usepackage{epsf}

\usepackage{mathtools}

\usepackage{amsfonts}
\usepackage{algorithmic}

\usepackage{textcomp}

\usepackage{textcomp}
\usepackage{setspace}
\usepackage{psfrag}
\usepackage{color,soul}
\usepackage{algorithm}
\usepackage{bbold}
\usepackage{verbatim}
\usepackage{amsthm}
\usepackage{braket}
\usepackage{enumitem}
\usepackage{lettrine}
\usepackage{wrapfig}
\usepackage{times}
\usepackage[english]{babel}
\usepackage{mathrsfs}  
\usepackage{amssymb,  amsmath,  appendix}
\usepackage{tikz}

\newtheorem*{lemma*}{Lemma}

\newtheorem{theorem}{Theorem}
\newtheorem*{theorem*}{Theorem}

\theoremstyle{definition}

\newcommand{\abs}[1]{\left\vert#1\right\vert}
\newcommand{\norm}[1]{\left\lVert #1\right\rVert}
\newcommand{\tr}[1]{ \mathrm{tr}\left(#1\right) }

\newcommand{\defeq}{\vcentcolon=}

\DeclareMathOperator{\vect}{vec}
\DeclareMathOperator{\diag}{diag}
\DeclareMathOperator{\rank}{rank}
\DeclareMathOperator{\spn}{span}
\DeclareMathOperator{\Ima}{Im}

\begin{document}


\title{Correlation Minor Norms, Entanglement Detection and Discord}

\author{Bar Y. Peled}
\thanks{These two authors contributed equally}
\affiliation{Center for Quantum Information Science and Technology \& Faculty
	of Engineering Sciences, Ben-Gurion University of the Negev, Beersheba
	8410501, Israel}
\author{Amit Te'eni}%
\thanks{These two authors contributed equally}
\affiliation{Faculty of Engineering and the Institute of Nanotechnology and Advanced
Materials, Bar Ilan University, Ramat Gan 5290002, Israel}
\author{Avishy Carmi}
\affiliation{Center for Quantum Information Science and Technology \& Faculty
	of Engineering Sciences, Ben-Gurion University of the Negev, Beersheba
	8410501, Israel}
\author{Eliahu Cohen}
\affiliation{Faculty of Engineering and the Institute of Nanotechnology and Advanced
	Materials, Bar Ilan University, Ramat Gan 5290002, Israel}

\date{\today}

\begin{abstract}
In this paper we develop an approach for detecting entanglement, which is based on measuring quantum correlations and constructing a correlation matrix. The correlation matrix is then used for defining a family of parameters, named Correlation Minor Norms, which allow one to detect entanglement.
This approach generalizes the computable cross-norm or realignment (CCNR) criterion, and moreover requires measuring a state-independent set of operators.
Furthermore, we illustrate a scheme which yields for each Correlation Minor Norm a separable state that maximizes it. The proposed entanglement detection scheme is believed to be advantageous in comparison to other methods because correlations have a simple, intuitive meaning and in addition they can be directly measured in experiment. Moreover, it is demonstrated to be stronger than the CCNR criterion. We also illustrate the relation between the Correlation Minor Norm and entanglement entropy for pure states. Finally, we discuss the relation between the Correlation Minor Norm and quantum discord. We demonstrate that the CMN may be used to define a new measure for quantum discord.
\end{abstract}

\pacs{Valid PACS appear here}
\maketitle


\section{Introduction}
The last three decades have seen significant advancement in development of promising quantum technologies, both from theoretical and practical aspects. These technologies often utilize quantum entanglement in order to gain advantage compared to classical technologies. Thus, the practical ability to detect entanglement 
is essential for the advancement of quantum technologies. Entanglement detection in many-body quantum systems is also of major interest~\cite{islam2015measuring,amico2008entanglement,jurcevic2014quasiparticle,kaufman2016quantum}, as well as quantum correlations in various physical settings such as those occurring in quantum optics~\cite{braunstein2005quantum,bello2020complex,berrada2011entanglement,abdel2012beam}, solid-state physics~\cite{chtchelkatchev2002bell,wiesniak2005magnetic,gonzalez2013mesoscopic} and atomic physics~\cite{tichy2011essential,sackett2000experimental,jaksch1999entanglement,yonacc2006sudden,berrada2012quantum,mohamed2019non}.

This has led researchers to seek simple ways to detect entanglement, preferably, ones which may be used in practice. For example, the Peres-Horodecki criterion~\cite{peres1996separability} is a necessary condition for a state to be separable; however, it is sufficient only in the $2 \times 2$ and $2 \times 3$ dimensional cases~\cite{horodeckis1996separability,horodecki1997separability}.

Another important concept is an entanglement witness, which is a measurable quantum property (i.e. a bounded Hermitian operator), such that its expectation value is always non-negative for separable states~\cite{horodeckis1996separability}. For any entangled state, there is at least one entanglement witness which would achieve a negative expectation value in this state. Alas, to use an entanglement witness in order to detect entanglement, one must measure a specific operator tailored to the state. An approach to quantify entanglement using entanglement witnesses can be found in~\cite{brandao2005quantifying}.

In \cite{guhne2004characterizing,guhne2007covariance,guhne2006entanglement,gittsovich2008unifying,li2018necessary,vicente2007separability}, a construction of a quantum correlation matrix was demonstrated, and it was shown that this matrix may be utilized to detect entanglement. In \cite{carmi2018significance,carmi2019relativistic}, a quantum correlation matrix has allowed the authors to derive generalized uncertainty relations, as well as a novel approach for finding bounds on nonlocal correlations. This matrix is the correlation matrix of a vector of quantum observables; thus, it may have complex entries. In~\cite{te2019multiplicative} it was demonstrated that such a matrix allows one to construct new Bell parameters and find their Tsirelson bounds. Another approach for Bell parameters based on covariance can be found in~\cite{Pozsgay}.

Indeed, quantum correlations are subtly related to entanglement, e.g. pure product states are always uncorrelated. This is not true for mixed states: separable mixed states may admit quantum correlations between remote parties~\cite{ollivier2001quantum}. These correlations are due to noncommutativity of quantum operators; hence, they allude to a different quantum property aside of entanglement, known as quantum discord~\cite{zurek2000einselection,ollivier2001quantum,henderson2001classical,giorda2010gaussian,luo2008quantum,bera2017quantum}. Since quantum discord is generally hard to compute when using its original definition, researchers have examined other discord measures which are more computationally tractable - most notably, geometric quantum discord \cite{dakic2010necessary,luo2010geometric}.

In~\cite{lupo2008bipartite,li2011note}, an approach for detecting entanglement using symmetric polynomials of the state's Schimdt coefficients has been studied. It was shown to be a generalization of the well-known CCNR criterion (computable cross-norm or realignment; first defined in~\cite{chen2002matrix,rudolph2005further}), according to which the sum of all Schmidt coefficients is no greater than $1$ for any separable state. The symmetric polynomial approach equips each one of these polynomials with some upper bound, and if the polynomial exceeds its bound then it follows that the state is entangled. Therefore, the sum of all Schmidt coefficients with the upper bound $1$ is a special case of this approach.

In this paper, we construct for a given quantum state its quantum correlation matrix, and examine the norms of its \textit{compound} matrices. Since the compound matrix in our case is constructed from minors of a certain correlation matrix, we call the proposed entanglement detectors ``Correlation Minor Norms''.
Seeing that these norms are invariant under orthogonal transformations of the observables, they can be regarded as a family of physical scalars which can be readily derived from bipartite correlations. Next, for each Correlation Minor Norm (CMN) we find an upper bound, such that if the CMN exceeds this bound it is implied that the state is entangled. Our proposed method is shown to generalize the symmetric polynomial approach. We also provide results and conjectures regarding the states that saturate the bounds. Moreover, we explore how the CMN relates to entanglement entropy.
Next, we construct a novel measure for quantum discord based on the CMN. In a particular case, it is identical to geometric quantum discord. We conclude by discussing possible generalizations for multipartite scenarios.

\section{Construction of the correlation matrix}
Let two remote parties, Alice and Bob, share a quantum system in $\mathcal{H}_A \otimes \mathcal{H}_B$, the tensor product of Hilbert spaces. Denote $d_A \defeq \dim \mathcal{H}_A , d_B \defeq \dim \mathcal{H}_B $, and let $ \bm{A} \defeq \left\{ A_i \right\}_{i=1}^{d_A^2}  $ be an orthonormal basis of the (real) vector space of $ d_A \times d_A $ Hermitian operators, w.r.t. the Hilbert-Schmidt inner product. Similarly, $ \bm{B} \defeq \left\{ B_j \right\}_{j=1}^{d_B^2} $ is an orthonormal basis of the $ d_B \times d_B $ Hermitian matrices. Note that such a basis always exists, since the real vector space of $ n \times n $ Hermitian matrices is simply the real Lie algebra $ \mathfrak{u} \left( n \right) $, which is known to have dimension $ n^2 $. Here we regard $ \mathfrak{u} \left( d_{A} \right), \mathfrak{u} \left( d_{B} \right) $ simply as inner product spaces, ignoring their Lie algebraic properties. Consequentially, we require the normalization $\tr{A_i A_j} = \delta_{ij} $ (and similarly for Bob) - without the factor of $2$, which is normally taken to make the structure constants more convenient. For example, for $d = 3$ one could take $ A_9 = \frac{1}{\sqrt{3}} \mathbb{1} $ and $ A_i = \frac{1}{\sqrt{2}} \gamma_i $ for all $i=1, \ldots, 8$, where $\gamma_i$ are the (``standard-normalization'') Gell-Mann matrices.

The (cross-)correlation matrix of $ \bm{A}, \bm{B} $, denoted by $\mathcal{C}$, is defined by:
\begin{align}\label{cor_ij}
	\mathcal{C}_{ij} & \defeq \braket{ A_i \otimes B_j } = \tr{ \rho A_i \otimes B_j}
\end{align}
where $\rho$ is the density matrix shared by Alice and Bob.

As we shall see in the next section, the information contained in $ \mathcal{C} $ regarding the strength of nonlocal correlations is encoded entirely in its singular values.
An equivalent characterization is provided by a noteworthy relation between the singular value decomposition (SVD) of $ \mathcal{C} $ and the \textit{operator-Schmidt decomposition} of the underlying state, which we describe hereinafter. The operator-Schmidt decomposition of any state $ \rho $ is defined as its unique decomposition of the form:
\begin{equation}
	\rho = \sum_{k=1}^{d^2} \lambda_k G_k \otimes H_k
\end{equation}
where each $ \lambda_k \geq 0 $ is a real scalar, and the sets $ \left\{ G_k \right\} $ and $ \left\{ H_k \right\} $ are orthonormal sets of $ d_{A} \times d_{A}, d_{B} \times d_{B} $ Hermitian matrices respectively. It can be shown that the singular values of $ \mathcal{C} $ are precisely the Schmidt coefficients $ \lambda_k $; moreover, the sets $ \left\{ G_k \right\} $ and $ \left\{ H_k \right\} $ are related to the sets $ \left\{ A_i \right\} $ and $ \left\{ B_j \right\} $ through the orthogonal matrices $U,V$ of the SVD, respectively.
Extended definitions and proof may be found in Appendix \ref{app_sub:operator_Schmidt}.

\section{Correlation Minor Norm}
The goal of this work is to produce physical scalars from $ \mathcal{C} $ that would allow for entanglement detection. In the context of this paper, a scalar is considered to be physical if it is invariant under a transformation of the set of measurements. Such a transformation is described by a pair of orthogonal matrices (see discussion in Appendix \ref{app_sub_basis}):

\begin{equation}\label{ortho_transform}
	\mathcal{C} \rightarrow \mathcal{U}_A C \mathcal{U}_B^T , \qquad \mathcal{U}_{A} \in \mathrm{O} \left( d_{A}^2 \right) , \mathcal{U}_{B} \in \mathrm{O} \left( d_{B}^2 \right) .
\end{equation}
Introducing into \eqref{ortho_transform} the SVD of $ \mathcal{C} $, written as $ \mathcal{C} = \mathcal{V}_A \Sigma \mathcal{V_B}^T $, yields:
\begin{equation}\label{ortho_transform_SVD}
	\mathcal{V}_A \Sigma \mathcal{V}_B^T \rightarrow \mathcal{U}_A \mathcal{V}_A \Sigma \mathcal{V}_B^T \mathcal{U}_B^T .
\end{equation}
Since $ \mathcal{U}_{A}$ and $  \mathcal{V}_{A} $ are elements of $ \mathrm{O} \left( d_{A}^2 \right) $ (and similarly for the matrices with the subscript $B$), we may observe that a general orthogonal transformation of $ \mathcal{C} = \mathcal{V}_A \Sigma \mathcal{V_B}^T $ reduces to the substitution of $ \mathcal{V}_A $ and $ \mathcal{V}_B $ by any other elements of their respective orthogonal groups. Thus, it is clear that any physical scalar derived from $ \mathcal{C} $ should depend on its singular values, i.e. the operator-Schmidt coefficients.

The simplest candidates for scalars produced by a matrix are its trace, determinant, and any type of matrix norm. However, $ \tr{\mathcal{C}} $ is not a physical scalar in the sense described above; and a broad class of matrix norms are given as special cases of the scalars constructed in this section. Thus, for now we wish to consider $ \det \mathcal{C} $ (where the discussion is restricted to $d_A = d_B$). The determinant of a quantum cross-correlation matrix can help detect entanglement, and may also serve as a measure of entanglement for two-qubit pure states (see Appendix \ref{app_two_qubit_pure}) and two-mode Gaussian states~\cite{simon2000peres,dodonov2004separability,de2006purity}.

However, in more general scenarios, there are states in which the mutual information between Alice and Bob stems from specific subspaces of their respective vector spaces (in pure states, the dimension of these subspaces is given by the Schmidt rank). To accommodate these cases, one should go over all possible subspaces of some given dimension and consider the determinant of the matrix comprised of correlations between their basis elements. Then, one could construct a measure as some function of all those determinants. One way of doing so is treat them as entries of a matrix and take its \textit{norm}.

In light of the observations above, we define the \textit{Correlation Minor Norm with parameters $h$ and $p=2$}:
\begin{equation}\label{M_h_2}
	\mathcal{M}_{h,p=2} \defeq \sqrt{ \sum_{R \in \binom{ \left[ d_A^2 \right] }{ h } } \sum_{S \in \binom{ \left[ d_B^2 \right] }{ h } } \abs{ \det \mathcal{C}_{R, S } }^2 }
\end{equation}
where $\binom{ \left[ a \right] }{b}$ denotes the set of $b$-combinations of $ \left[a\right]$ (this notation is common in the Cauchy-Binet formula), $\mathcal{C}_{R,S}$ is the matrix whose rows are the rows of $\mathcal{C}$ at indices from $R$ and whose columns are the columns of $\mathcal{C}$ at indices from $S$, and $1 \leq h \leq \min \left\{ d_A^2, d_B^2 \right\} $. The meaning of the parameter $p$ will become clear shortly, when the above definition is generalized.

Note that $ \mathcal{M}_{h,2} $ is the Frobenius norm of a matrix $\mathcal{N}$ of size $ \binom{ d_A^2 }{ h } \times \binom{ d_B^2 }{ h } $, defined by:
\begin{equation}
	\mathcal{N}_{ij} \triangleq \det \mathcal{C}_{R_i, S_j}
\end{equation}
where we have numbered the sets' elements:
\begin{align*}
	& \binom{ \left[ d_A^2 \right] }{ h } \defeq \left\{ R_1, \ldots, R_{ \binom{ d_A^2 }{ h } } \right\} , \\
	& \binom{ \left[ d_B^2 \right] }{ h } \defeq \left\{ S_1, \ldots, S_{ \binom{ d_B^2 }{ h } } \right\} .
\end{align*}
Such a matrix $ \mathcal{N} $ is known as the \textit{$h$-th compound matrix of $ \mathcal{C}$}, and is denoted by $C_h \left( \mathcal{C} \right)$. Now, recall the Schatten $p$-norm of any matrix $M$ is defined by $ \norm{M}_p \defeq \norm{ \vec{\sigma} \left( M \right) }_p $, i.e. the \textit{vector} $p$-norm of the vector composed of the singular values of $ M $. Schatten $p$-norms lead to a generalization of the definition \eqref{M_h_2}: for $ p \in \left[ 1 , \infty \right)$, define the \textit{Correlation Minor Norm with parameters $h$ and $p$} as:
\begin{equation}
	\mathcal{M}_{h,p} = \norm{ C_h \left( \mathcal{C} \right) }_p ,
\end{equation}
i.e., it is the Schatten $p$-norm of the $h$-th compound matrix of the correlation matrix $ \mathcal{C} $.
Substituting the known relation between the singular values of any matrix and its compound matrix (see Appendix \ref{app_CMN_SVD}), one obtains the following formula for computing the Correlation Minor Norm (CMN):
\begin{equation}\label{M_h_p}
	\mathcal{M}_{h,p} = \left( \sum_{R \in \binom{ \left[ d^2 \right] }{h} } \prod_{k \in R} \left[ \sigma_k \left( \mathcal{C} \right) \right]^p \right)^{1/p} ,
\end{equation}
where $ d = \min \left\{ d_A, d_B \right\} $, and $ \sigma_k \left( \mathcal{C} \right) $ denotes the $k$-th singular value of $\mathcal{C}$. This implies that $\mathcal{M}_{h,p}$ is indeed a physical scalar. Note that the Schatten $p$-norm of $\mathcal{C}$ itself is obtained as a special case, for $ h=1 $. Another thing to note is that for $h = d^2$, the CMN $ \mathcal{M}_{h,p} $ is equal to the product of all singular values, irregardless of $p$; in this case we denote it by $ \mathcal{M}_{h=d^2} $. If $d_A=d_B$, this is simply $\det \mathcal{C}$.

\section{Entanglement detection using the Correlation Minor Norm}
For general mixed states, there are a few known links between Schmidt coefficients and entanglement detection; the best-known is probably the CCNR criterion: If $ \sum_{k=1}^{d^2} \lambda_k > 1$, then $\rho$ is entangled~\cite{guhne2009entanglement}. The Correlation Minor Norm allows for an equivalent formulation: if $ \mathcal{M}_{h=1,p=1} >1 $, then $\rho$ is entangled.

The CCNR criterion has an additional immediate consequence regarding the CMN: since $ \mathcal{M}_{h,p} $ is a monotonically increasing function of the operator-Schmidt coefficients $ \lambda_k $, there is an upper bound for the value it may obtain without violating the inequality $ \sum_{k=1}^{d^2} \lambda_k \leq 1$. Thus, for all $h$ and $p$, there exists some positive number $B = B \left( d_A, d_B, h, p \right)$ with the property: if $\rho$ is separable, then $ \mathcal{M}_{h,p} \leq B \left( d_A, d_B, h, p \right) $. This implies the Correlation Minor Norm can be used to detect entanglement by the following procedure: given a state $ \rho $, the corresponding correlation matrix $ \mathcal{C} $ is obtained - either by computation or by direct measurement; then, the SVD of $ \mathcal{C} $ is used to find the singular values, and these are substituted in \eqref{M_h_p} to compute the desired CMN, $ \mathcal{M}_{h,p} $; and finally, $ \mathcal{M}_{h,p} $ is compared with $ B \left( d_A, d_B, h, p \right) $. If $ \mathcal{M}_{h,p} \leq B \left( d_A, d_B, h, p \right) $, we cannot deduce anything. However, if $ \mathcal{M}_{h,p} > B \left( d_A, d_B, h, p \right) $, we infer the state $ \rho $ is entangled. The remainder of this section deals with results regarding the upper bounds $ B \left( d_A, d_B, h, p \right) $. A technical treatment of the operator-Schmidt decomposition for separable states appears in Appendix \ref{operator_Schmidt_separable}.

In \cite{lupo2008bipartite}, Lupo et al. generalize the CCNR criterion in the following way: they construct all \textit{elementary symmetric polynomials} of the Schmidt coefficients $\lambda_k$ of $\rho$, and find bounds on these assuming $\rho$ is separable. The $h$-th elementary symmetric polynomial of $n$ variables is defined as follows:
\begin{equation}
	S_h \left( x_1, \ldots, x_n \right) \defeq \sum_{R \in \binom{[n]}{h}} \prod_{k \in R} x_k ,
\end{equation}
i.e., the sum of all distinct products of $h$ distinct variables. Clearly, $ \mathcal{ M}_{h,p=1} = S_h \left( \sigma_1, \ldots, \sigma_{d^2} \right) $.

A more recent work~\cite{li2011note} which cites \cite{lupo2008bipartite}, makes the following important claim: assuming $d_A = d_B$, they find a tight bound on the $h$-th symmetric polynomial (for separable states), and prove that as an entanglement detector it is no stronger than the CCNR criterion. Since the conjectures presented in this section imply this is true for the CMN with $ p = \infty $ as well, it seems likely that for $ d_A = d_B $ and any value of $ p $, the CMN is no stronger than the CCNR criterion as an entanglement detector.

However, in the case where $d_A \neq d_B$, it seems the CMN may detect entanglement in cases where CCNR does not. Let us define following~\cite{kent1999optimal,verstraete2003normal,leinaas2006geometrical}, \textit{a state in Filter Normal Form (FNF)} as a state $ \rho $ for which any traceless Alice-observable $A$ and any traceless Bob-observable $B$ have vanishing expectation values; i.e., $ \braket{A \otimes \mathbb{1}}_\rho = \braket{\mathbb{1} \otimes B}_\rho = 0 $. Then, we have the following result:
\begin{theorem}\label{thm_bound_p_1}
	Assume $D \defeq \max \left\{ d_A, d_B \right\} \leq d^3 $ and $h >1$. Then, for any separable state in Filter Normal Form:
	\begin{equation}\label{bound_p_1}
		\mathcal{M}_{h,p=1} \leq S_h \left( \alpha, \beta, \ldots, \beta \right)
	\end{equation}
	where $ \alpha \defeq 1/\sqrt{Dd} $, $ \beta \defeq \sqrt{ \frac{D-1}{D \left(d^2-1\right)} \frac{d-1}{d \left(d^2-1\right)} } $, and $S_h$ is the $h$-th elementary symmetric polynomial in $d^2$ variables.
\end{theorem}
Proof may be found in Appendix \ref{app_proof_thm_1}. Moreover, we conjecture the following theorem still holds with the assumption of the state being in FNF removed. If proven, this conjecture would have explained the upper bounds presented in \cite{lupo2008bipartite} for $d_A \neq d_B$, which had been found numerically.

Before presenting the next result, let us introduce quantum designs~\cite{zauner2011quantum}. A \textit{quantum design} in dimension $b$ with $ v $ elements is simply a set of $v$ orthogonal projections $ \left\{ P_k \right\}_{k=1}^v $ on $ \mathbb{C}^b $. A quantum design is \textit{regular with $r=1$} if all projections are pure (i.e. one-dimensional); it is \textit{coherent} if the sum $ \sum_k P_k $ is proportional to the identity operator; and it has \textit{degree $1$} if there exists $ \mu \in \mathbb{R} $ such that $ \forall k \neq l, \tr{P_k P_l} = \mu $. If a quantum design has all three qualities, then $ \mu = \frac{v  -b }{b \left( v-1 \right)} $.

A regular, coherent, degree-$1$ quantum design with $r = 1$ having $ v $ elements, is simply a set of $ v $ ``equally spaced'' pure states in the same space. For example, such a quantum design in dimension $d$ containing $d^2$ elements is known as a symmetric, informationally complete, positive operator-valued measure (SIC-POVM)~\cite{renes2004symmetric}. 

The following theorem tells us how to construct a separable state saturating \eqref{bound_p_1} using quantum designs.
\begin{theorem}\label{maximizing_p_1}
	Let $ \left\{ P_k^A \right\}_{k=1}^{d^2} $, $ \left\{ P_k^B \right\}_{k=1}^{d^2} $ be sets of pure projections comprising regular, coherent, degree-$1$ quantum designs with $r = 1$, in dimensions $d_A, d_B$ respectively and having $d^2$ elements each. Define a state:
	\begin{equation}\label{rho_designs}
		\rho = \frac{1}{d^2} \sum_{k=1}^{d^2} P_k^A \otimes P_k^B
	\end{equation}
	Then, the operator-Schmidt coefficients of $ \rho $ are $ \alpha $ with multiplicity one and $ \beta $ with multiplicity $d^2-1$.
\end{theorem}
The proof appears in Appendix \ref{sat_p_equals_1}.
Note the last two theorems have the following special case: for $h=d^2$, they imply that the above state maximizes the product of all Schmidt coefficients; i.e., it maximizes $ \mathcal{M}_{h=d^2,p} $ for all $ p $.

Furthermore, we have similar claims for $p=\infty$.
\begin{theorem}\label{thm_bound_p_infty}
	Let $ \rho $ be a separable state in FNF, and $h \geq \sqrt{Dd}$. Then:
	\begin{equation}\label{bound_p_infty}
		\mathcal{M}_{h,p=\infty} \leq \frac{1}{\sqrt{Dd}} \left[ \frac{D-1}{D \left(h-1\right)} \frac{d-1}{d \left(h-1\right)} \right]^{\frac{h-1}{2}} .
	\end{equation}
\end{theorem}
Proof may be found in Appendix \ref{app_proof_thm_3}. As in Theorem \ref{thm_bound_p_1}, we conjecture this theorem still holds without the assumption that $\rho$ is in FNF. Evidence for why we believe this conjecture to be true may be found in Appendix \ref{app_evidence}. The following theorem yields a way of saturating the bound \eqref{bound_p_infty}:
\begin{theorem}\label{maximizing_p_infty}
	Let $ \left\{ P_k^A \right\}_{k=1}^{h} $, $ \left\{ P_k^B \right\}_{k=1}^{h} $ be sets of pure projections comprising regular, coherent, degree-$1$ quantum designs with $r = 1$, in dimensions $d_A, d_B$ respectively and having $h$ elements each. Define a state:
	\begin{equation}\label{rho_h_designs}
		\rho = \frac{1}{h} \sum_{k=1}^{h} P_k^A \otimes P_k^B
	\end{equation}
	Then, the operator-Schmidt coefficients of $ \rho $ are $ \alpha $ with multiplicity one and $ \beta' = \sqrt{ \frac{D-1}{D \left(h-1\right)} \frac{d-1}{d \left(h-1\right)} } $ with multiplicity $h-1$.
\end{theorem}
The proof appears in Appendix \ref{app_proof_thm_4}. Note the coherence of $ \left\{ P_k^{A/B} \right\} $ ensures the state \eqref{rho_h_designs} is in FNF. Moreover, the constants $ \mu_{A/B} \defeq \frac{h -d_{A/B} }{d_{A/B} \left( h-1 \right) } $ enter the operator-Schmidt coefficients (and thus the upper bound \eqref{bound_p_infty}) elegantly: $ \beta' = \sqrt{ \frac{1-\mu_A}{h} \frac{1-\mu_B}{h} } $.

We hypothesize that upper bounds over $\mathcal{M}_{h,p}$ for any value of $p$ may be characterized using quantum designs. If this hypothesis is proven, then separable states built using such quantum designs are, in a way, on the ``edges'' of the convex separable set. However, one should note that quantum designs in a given dimension with a given number of elements do not always exist; the above theorems hold only in the cases where they do exist.

\section{Further results and open questions}

\subsection{Relation to entanglement entropy for pure states}
Let $\ket{\psi}$ be a pure state, and let $s_1,\ldots,s_d$ denote its ``pure-state-Schmidt coefficients'' (i.e., the ones arising when writing the Schmidt decomposition for pure states of $\ket{\psi}$). Then, its operator-Schmidt coefficients are $ s_k s_l $, i.e. all the pairwise products of pure-state-Schmidt coefficients (if $k \neq l$, $ s_k s_l $ appears as an operator-Schmidt coefficient with multiplicity $2$; for proof please refer to Appendix \ref{app_operator_Schmidt_pure}.

\begin{figure}
	\centering
	
	\includegraphics[width=\linewidth]{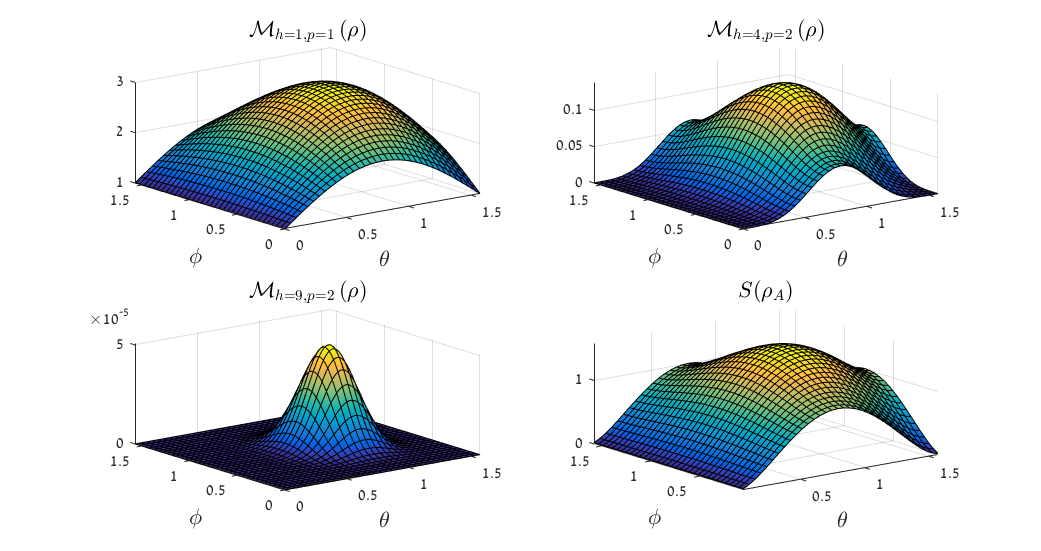} 
	
	\caption{Several CMNs $ \mathcal{M}_{h,p} $ and the von-Neumann entanglement entropy $S \left( \rho_A \right)$, plotted for the two-qutrit states with pure-state-Schmidt coefficients given in spherical coordinates: $s_1 = \sin \theta \cos \phi$, $s_2 = \sin \theta \sin \phi $ and $ s_3 = \cos \theta $ (to avoid repetition, only the domain $ 0 \leq \theta, \pi \leq \pi /2 $ is plotted). The area where $ S \left( \rho_A \right) $ vanishes, $ \theta = 0, \pi / 2 $, alludes to the domain where the state's Schmidt rank is $1$, and the same goes for $ \mathcal{M}_{h=2^2,p} $. However, $ \mathcal{M}_{h=3^2,p} $ also tells us where the Schmidt rank is $2$, i.e., $ \phi = 0, \pi/2 $. (This figure was created using MATLAB R2016A)}
	\label{fig:vNvsM1}
\end{figure}

For pure states, the Correlation Minor Norm is linked to the state's Schmidt rank by the following observation: for all $t \in \left[ d \right]$, $\mathcal{M}_{h=t^2,p} \neq 0$ iff the state's pure-state-Schmidt rank is at least $t$. Thus, the Correlation Minor Norm may be used to find the Schmidt rank in pure states. Fig. \ref{fig:vNvsM1} illustrates a comparison between $\mathcal{M}_{h=t^2, p}$ and entanglement entropy for all two-qutrit pure states.

Moreover, for any pure state of dimension $2 \times D$, the CMN and entanglement entropy are only functions of $s_1^2$, and both functions have the same monotonicity w.r.t. this parameter (i.e. they increase / decrease in the same domains). This may be demonstrated by noting that effectively, the qubit is only correlated with a two-dimensional subsystem of Bob's system. Using the same reasoning that appears in Appendix \ref{app_two_qubit_pure}, it could be argued that $\mathcal{M}_{h=4} $ may indeed quantify entanglement in this scenario.

However, it is clear that not all Correlation Minor Norms are useful for this purpose; in fact, the relation between operator-Schmidt coefficients and pure-state-Schmidt coefficients implies:
\begin{equation}
	\mathcal{M}_{h=1,p=2}^2 = \sum_k \lambda_k^2 = \sum_{k,l} \left( s_k s_l \right)^2 = \left( \sum_k s_k^2 \right)^2 = 1 .
\end{equation}
Thus, $ \mathcal{M}_{h=1,p=2} = 1 $ for \textit{any} pure state, be it separable or entangled.

\subsection{Improving on the CCNR criterion}
In this section, we shall present an entangled state which may be detected by the CMN, but \textit{cannot} be detected by the CCNR criterion. First, let $ \rho_0 $ be the state \eqref{rho_designs} for $ d_A = 3, d_B = 2 $; and let $ \rho_1 \defeq \ket{\psi} \bra{\psi} $, where $ \psi \defeq \left( \ket{11} + \ket{20} \right)/\sqrt{2} $. The state is constructed as follows:
\begin{equation}
	\rho_q = q \rho_1 + \left(1-q\right) \rho_0 .
\end{equation}
For $ q=0.295 $ the state is entangled (easily verifiable by the PPT criterion). However, it is not detected by the CCNR criterion: $ \mathcal{M}_{h=1,p=1} = 0.9981 < 1 $; and it is detected by the CMN: $ \mathcal{M}_{h=2,p=1} = 0.3509 $, exceeding the bound $ \frac{2+ 3 \sqrt{2} }{18} \approx 0.3468 $.

\subsection{Relation to quantum discord}
Since the CMN seems to capture some value related to quantum correlations, it is intriguing to ask whether it may somehow be be used to measure their strength. The geometric measure for quantum discord (GQD) with respect to Alice's subsystem is defined~\cite{dakic2010necessary} as:
\begin{equation}
	\mathcal{D}_G^A \left( \rho \right) \defeq \min_{\chi \in c-q} \norm{ \rho - \chi }^2 ,
\end{equation}
i.e. the shortest squared Euclidean distance between $\rho$ and any classical-quantum state (the expression for discord w.r.t. Bob's subsystem is defined similarly, where the minimization goes over all quantum-classical states).

Motivated by this definition and by the expression for GQD derived in~\cite{luo2010geometric}, we suggest the following measure for discord w.r.t. Alice's subsystem, based on the CMN:
\begin{equation}\label{Discord_CMN}
	\mathcal{D}^A_{h,p} \left( \rho \right) = \left[ \mathcal{M}_{h,p} \left( \rho \right) \right]^p  - \max_{\Pi^A \in M \left( A \right) } \left[ \mathcal{M}_{h,p} \left( \Pi^A \left[ \rho \right] \right) \right]^p ,
\end{equation}
where the maximization goes over all projective measurements on Alice's subsystem $ \Pi^A = \left\{ \Pi_i \right\}_{i=1}^{d_A} $, and $ \Pi^A \left[ \rho \right] $ is the state obtained from $\rho$ by performing the measurement $ \Pi^A $ and obtaining the appropriate ensemble of the projections $\Pi_i$ (i.e., the state is measured but not ``collapsed''). The following result suggests that $ \mathcal{D}^A_{h,p} $ may be thought of as a measure for discord:
\begin{theorem}
	For any state $\rho$ and for any value of $h, p$, $ \mathcal{D}^A_{h,p} \left( \rho \right) \geq 0 $; and $ \mathcal{D}^A_{h\leq 2,p} \left( \rho \right) = 0 $ iff $ \mathcal{D}_G^A \left( \rho \right) = 0 $.
\end{theorem}

Moreover, for any state $\rho$, we have $ \mathcal{D}_G^A \left( \rho \right) = \mathcal{D}^A_{h=1,p=2}$. The proof for this fact, as well as for the theorem above, appears in Appendix \ref{app_discord}. As evident in the proof, $ \mathcal{D}_G^A \left( \rho \right) = 0 \; \Rightarrow \; \mathcal{D}^A_{h,p} \left( \rho \right) = 0 $ for $ h>2 $ as well.

\begin{figure}
	\centering
	
	\includegraphics[width=\linewidth]{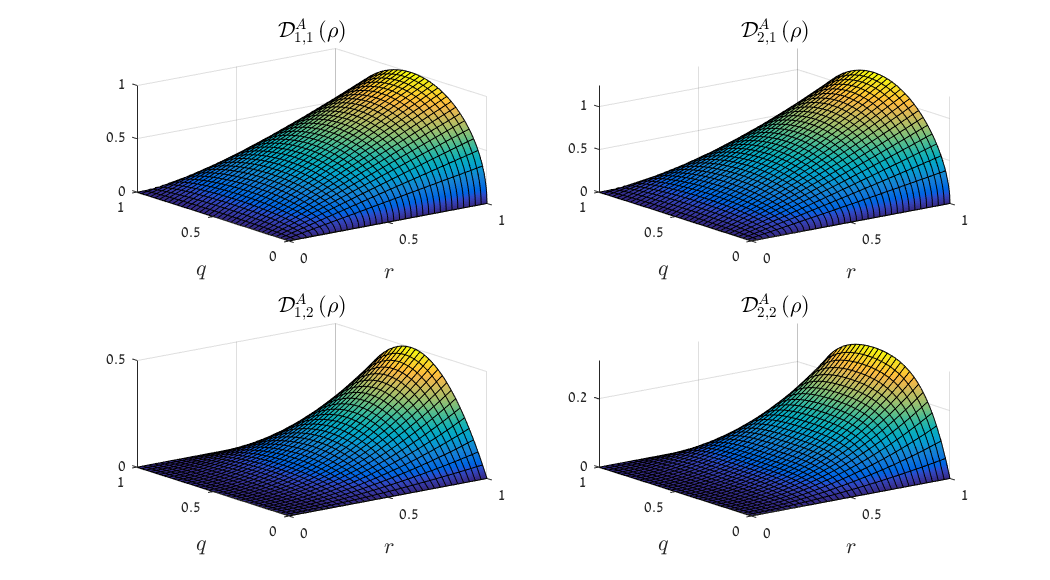}
	\caption{The CMN-motivated measures for geometric quantum discord $ \mathcal{D}_{h,p}^A $, for the family of states given in \cite{virzi2019optimal}, with parameters $q,r \in \left[ 0, 1 \right] $. The discord is computed using the formula given on \cite{virzi2019optimal}. The similarity is not a coincidence, as the geometric discord of these states is given by a product of two operator-Schmidt coefficients (times a factor of $2$). (This figure was created using MATLAB R2016A)}
	\label{fig:dis2}
\end{figure}

Figure \ref{fig:dis2} illustrates several of the measures $\mathcal{D}^A_{h,p} \left( \rho \right)$ for a two-parameter family of states given in \cite{virzi2019optimal}.
The states appear in Appendix \ref{app_discord}. It is also worth noting in this context, that the two-qubit separable state with maximal discord has the same operator-Schmidt coefficients as the Werner state with $c=1/3$~\cite{galve2011maximally,bera2017quantum} (hence they are unitarily equivalent); and it is precisely the state maximizing the CMNs with $p=1$. In other words - the entanglement, discord and CMNs for the two-qubit Werner states are all monotonically increasing functions of the parameter $c \in \left[ 0, 1 \right]$, and for the critical value of $c$ (above which the states are entangled) the Werner state is precisely the one for which the CMN obtains its separable upper bound. This situation occurs for the two-qutrit Werner state as well, where the critical value of $c$ is $1/4$ (when constructed as in \cite{ye2013analytic}).

\section{Conclusions}
The task of entanglement detection is important for basic quantum science, as well as various quantum technologies. The current work was motivated by the following question: since bipartite entanglement can be characterized by correlations between all of the parties' observables, can it also be detected via some norm of these correlations? As demonstrated by our results, the answer is likely to be affirmative.

We have defined the Correlation Minor Norm and explored its characteristics. This has allowed us to propose an approach for detecting entanglement both in pure and mixed states.
Furthermore, it was shown that for pure states, the Correlation Minor Norm allows one to determine the Schmidt rank, and in some cases also quantify the strength of quantum correlations. Given the dimensions of the two parties' respective systems, one may choose a single set of operators which can be used for detecting entanglement in any state, be it pure or mixed.

Additionally, we have shown that the CMN with $h=1,p=2$ admits a natural relation to geometric quantum discord. This affinity motivated a definition of a more general measure for quantum discord which is based on the CMN. Some of these measures might mitigate the known issues with existing discord measures \cite{piani2012problem,tufarelli2012quantum,paula2013geometric,roga2016geometric}.

One optional direction for future research may include development of dynamical equations for the Correlation Minor Norm. This may be interesting, as the correlation matrix contains exactly the same information as the density matrix.

Another possible generalization is considering multipartite systems.
In~\cite{de2011multipartite}, the authors consider detection of genuine multipartite entanglement and non-full-separability using correlation tensors. Specifically, they consider tensors comprising all multipartite correlations between orthonormal bases to the \textit{traceless} observables; and they find upper bounds on norms of \textit{matricizations} of these tensors, such that exceeding these bounds implies the state is genuine multipartite entangled, or non-fully-separable.

This paper may hint as to how our work may be generalized to the multipartite case: one could consider the \textit{full} correlation tensor (i.e. correlations between bases to the entire space of observables, not just the traceless ones); then, the CMN with parameters $h,p$ may be defined as the Schatten $p$-norm of the $h$th compound matrix of a certain matricization of this tensor. The bounds shown in~\cite{de2011multipartite} could then be utilized to find two upper bounds on each of the CMNs - one for non-genuinely-entangled states, and another for fully-separable states. The question of which matricization should be used remains to be determined. Moreover, further work is required to find the states saturating these bounds.

\section*{Acknowledgments}
We thank Aharon Brodutch for insightful discussions. B.P. and A.T. also thank Ebrahim Karimi for their hospitality at the University of Ottawa and Elie Wolfe for their hospitality in Perimeter Institute. Both visits have been fruitful and advanced this work. E.C. acknowledges support from the Israel Innovation Authority under project 70002, from FQXi (grant no. 224321), from the Pazy Foundation and from the Quantum Science and Technology Program of the Israeli Council of Higher Education.

\appendix

\section{Entanglement detection using the quantum correlation matrix}\label{app:quantum_corr_mat}

\subsection{The Operator-Schmidt Decomposition}\label{app_sub:operator_Schmidt}
Given any state $ \rho $ (either separable or entangled), one may write down the following unique decomposition:
\begin{equation}\label{Operator_Schmidt}
\rho = \sum_{k=1}^{d^2} \lambda_k G_k \otimes H_k
\end{equation}
where $d \defeq \min \left\{ d_A, d_B \right\}$, each $ \lambda_k \geq 0 $ is a real scalar, and the sets $ \left\{ G_k \right\} $ and $ \left\{ H_k \right\} $ form orthonormal bases of the $ d_{A/B} \times d_{A/B} $ Hermitian matrices. Note this is not necessarily a ``separable decomposition'', since $G_k, H_k $ are not compelled to be positive semi-definite. 

Let us assume that $ \lambda_k $ are in non-increasing order.
We shall demonstrate that the SVD of the cross-correlation matrix $ \mathcal{C} $ is equivalent to the Operator-Schmidt Decomposition. 
\begin{theorem*}\label{SVD_Schmidt}
	Given a state $\rho$, let $ \mathcal{C} $ be the second moment 
	matrix of the orthonormal sets $ \left\{ A_i \right\}, \left\{ B_j \right\} $, defined by $ \mathcal{C}_{ij} = \braket{A_i \otimes B_j}_\rho $. Let $ \mathcal{C} $ have the SVD $ \mathcal{C} = U \Sigma V^T $ with singular values $ \sigma_1 \geq \ldots \geq \sigma_{d^2} $. Then, the unique decomposition \eqref{Operator_Schmidt} of $\rho$ satisfies the following:
	\begin{enumerate}
		\item $ \lambda_k = \sigma_k $
		\item $ G_k = \sum_{i=1}^{d^2} U_{ik} A_i $
		\item $ H_k = \sum_{j=1}^{d^2} V_{jk} B_j $ .
	\end{enumerate}
\end{theorem*}
Proof outline: since $ \left\{ A_i \otimes B_j \right\}_{i,j} $ comprise a basis to the set of $d^2 \otimes d^2$ Hermitian matrices, the matrix $ \mathcal{C} $ suffices in order to fully characterize $\rho$. Thus, since the Operator-Schmidt decomposition is unique, all is left to do is verify that $ \rho $ with the above Operator-Schmidt decomposition reproduces the same correlations $ \mathcal{C}_{ij} $, which is straightforward. 

\subsection{Change of Measurement Basis}\label{app_sub_basis}
Mathematically, Alice and Bob's observables transform by the representation $\bm{d^2} = \bm{d^2-1 \oplus 1}$ (adjoint plus trivial singlet) of a local projective unitary transformation $ U \in \mathrm{PU} \left( d \right) $ (where $d$ is either $d_A$ or $d_B$):
\begin{equation}
    A_i \rightarrow U_A A_i U_A^\dagger , \; B_j \rightarrow U_B B_j U_B^\dagger ,
\end{equation}
where we have taken the projective unitary groups $\mathrm{PU} (d) = \mathrm{PSU} (d) = \mathrm{U} (d) / \mathrm{U} (1) = \mathrm{SU} (d) / \mathbb{Z}_d$, since any $ \mathrm{U} \left( 1 \right) $ phase clearly cancels out in the above. The trivial part is given by the identity component of $A_i$ (or $B_j$), and the adjoint part by the traceless component. Since the adjoint representation of $\mathrm{PU} (d)$ is a subgroup of $ \mathrm{SO} \left( d^2-1 \right) $, there exists a basis where the vector of observables $\left[ A_i \right]$ transforms by:
\begin{equation}\label{1_plus_SO}
    \begin{pmatrix}
    A_1 \\
    \vdots \\
    A_{d_A^2}
    \end{pmatrix} \rightarrow \begin{bmatrix}
    1 & 0 \\
    0 & R
    \end{bmatrix} \begin{pmatrix}
    A_1 \\
    \vdots \\
    A_{d_A^2}
    \end{pmatrix} ,
\end{equation}
where $A_1$ is a scalar matrix, and $ R \in \mathrm{SO} \left( d_A^2-1 \right) $.
For instance, suppose $d_A = d_B = 2$; then, Alice's observables may correspond to measurements of a spin-$1/2$ in a given orthonormal set of directions. If the first measurement is fixed to be the trivial one $ \mathbb{1} / \sqrt{2} $, then a special orthogonal transformation $ R \in \mathrm{SO} \left( 3 \right) $ describes a rotation of Alice's entire lab; similarly, Bob's lab may be rotated independently of Alice's.

However, if the first measurement is not fixed, then we should consider more general transformations than those with the form \eqref{1_plus_SO}. Since the required basis transformation preserves inner the Hilbert-Schmidt inner product, it can be taken to be a $d \times d$ orthogonal matrix; indeed, the matrices of the form \eqref{1_plus_SO} are naturally embedded in $\mathrm{O} (d_A^2)$.
Therefore, the correlation matrix furnishes a tensor product of two representations (i.e. Alice's and Bob's), described by
\begin{equation}
\mathcal{C} \rightarrow \mathcal{U}_A C \mathcal{U}_B^T , \qquad \mathcal{U}_{A} \in \mathrm{O} \left( d_{A}^2 \right) , \mathcal{U}_{B} \in \mathrm{O} \left( d_{B}^2 \right) .
\end{equation}

\subsection{Operator-Schmidt decomposition for pure states}\label{app_operator_Schmidt_pure}
Let $ \ket{\psi} $ be a pure state given in its pure-state-Schmidt decomposition:
\begin{equation}
\ket{\psi} = \sum_{k=1}^d s_k \ket{\phi_k} \otimes \ket{\xi_k} .
\end{equation}
The appropriate density matrix:
\begin{equation}\label{rho_pure_state}
\rho = \ket{\psi} \bra{\psi} = \sum_{k,l=1}^{d} s_k s_l \ket{\phi_k} \bra{\phi_l} \otimes \ket{\xi_k} \bra{\xi_l} .
\end{equation}
Let us fix $k, l$ s.t. $ k < l $. $k$ and $l$ appear in two terms of the sum: $ s_k s_l \left( \ket{\phi_k} \bra{\phi_l} \otimes \ket{\xi_k} \bra{\xi_l} + \ket{\phi_l} \bra{\phi_k} \otimes \ket{\xi_l} \bra{\xi_k} \right) $. We wish to write down the parenthesized expression in the form $ G_{kl} \otimes H_{kl} + G_{lk} \otimes H_{lk} $, where $ G_{kl}, H_{kl},  G_{lk}, H_{lk} $ are all trace-normalized Hermitian operators, and $ \tr{G_{kl} G_{lk}} = \tr{H_{kl} H_{lk}} = 0 $. Indeed, this is achieved by setting:
\begin{align}
& G_{kl} \defeq \frac{\ket{\phi_k} \bra{\phi_l} + \ket{\phi_l} \bra{\phi_k}}{\sqrt{2}} ; \nonumber\\ 
& H_{kl} \defeq \frac{\ket{\xi_k} \bra{\xi_l} + \ket{\xi_l} \bra{\xi_k}}{\sqrt{2}} ; \nonumber\\
& G_{lk} \defeq \frac{i \left( \ket{\phi_k} \bra{\phi_l} - \ket{\phi_l} \bra{\phi_k} \right)}{\sqrt{2}} ; \nonumber\\
& H_{lk} \defeq -\frac{i \left( \ket{\xi_k} \bra{\xi_l} - \ket{\xi_l} \bra{\xi_k} \right) }{\sqrt{2}} .
\end{align}
By supplementing the notations $ G_{kk} \defeq \ket{\phi_k} \bra{\phi_k} $, $ H_{kk} \defeq \ket{\xi_k} \bra{\xi_k} $, one may write \eqref{rho_pure_state} by:
\begin{equation}\label{operator_Schmidt_pure}
\rho = \sum_{k,l=1}^{d} s_k s_l G_{kl} \otimes H_{kl} .
\end{equation}
Since $ \left\{ G_{kl} \right\} $ and $ \left\{ H_{kl} \right\} $ are both orthonormal sets of operators, \eqref{operator_Schmidt_pure} is the operator-Schmidt decomposition of $ \rho $; thus, $ \left\{ s_k s_l \right\} $ are its operator-Schmidt coefficients.

\subsection{$\det \mathcal{C}$ for two-qubit pure states}\label{app_two_qubit_pure}
Let $ \ket{\psi} $ be a two-qubit pure state, given in its pure-state-Schmidt decomposition:
\begin{equation}
\ket{\psi} = s_1 \ket{\phi_1} \otimes \ket{\xi_1} + s_2 \ket{\phi_2} \otimes \ket{\xi_2} .
\end{equation}
From the previous subsection, its operator-Schmidt coefficients are $ s_1^2, s_1 s_2, s_2 s_1, s_2^2 $. Moreover, from Section \ref{app_sub:operator_Schmidt} of this supplemental material, these are also the singular values of its correlation matrix. Thus:
\begin{equation}
\det \mathcal{C} = \prod_k \sigma_k \left( \mathcal{C} \right) = s_1^4 s_2^4 = \left( s_1^2 s_2^2 \right)^2 = \left[ s_1^2 \left( 1- s_1^2 \right) \right]^2 ,
\end{equation}
where the final transition follows from the normalization condition $ s_1^2 + s_2^2 = 1 $. An interesting observation is that $ \sqrt[4]{\det \mathcal{C}} $ is proportional to the interferometric distinguishability measure studied in~\cite{jaeger1993complementarity,greenberger1988simultaneous,jaeger1995two,englert1996fringe,franson1989bell}; moreover, \cite{greenberger1988simultaneous} illustrates the striking resemblance between this measure and entangelement entropy. Thus, for two-qubit pure states, $ \det \mathcal{C} $ indeed quantifies entanglement. As an aside, we note that the distinguishability measure is generalized for a certain family of Gaussian states in~\cite{peled2020double}.

\section{CMN and SVD}\label{app_CMN_SVD}
In order to compute the CMN, one should seek a relation between the singular values of given matrix, and the singular values of its compound matrices. Such a relation is known~\cite{horn2012matrix}:
\begin{lemma*}
	Let $E$ be a $n \times n$ matrix. The singular values of $C_h \left( E \right) $, are the $\binom{n}{h}$ possible products $ \sigma_{i_1} \cdots \sigma_{i_h} $ .
\end{lemma*}
Which implies:
\begin{equation}\label{Frob_norm_of_compund_matrix}
\norm{ C_h \left( E \right) }_p = \left( \sum_{R \in \binom{ \left[ n \right] }{h} } \prod_{k \in R} \left[ \sigma_k \left( E \right) \right]^p \right)^{1/p}
\end{equation}
where: 
\begin{equation}
\binom{ \left[ n \right] }{h} \triangleq \left\{ R \in 2^{ \left[ n \right] } : \abs{R} = h \right\}
\end{equation}
i.e., $ \binom{ \left[ n \right] }{h} $ denotes the set of subsets of $ \left[ n \right] $ having cardinality $h$. Thus we obtain the following formula for the correlation minor norm, using only the singular values of the second moment matrix:
\begin{equation}\label{CMN_with_SVD}
\mathcal{M}_{h,p} = \left( \sum_{R \in \binom{ \left[ d^2 \right] }{h} } \prod_{k \in R} \left[ \sigma_k \left( \mathcal{C} \right) \right]^p \right)^{1/p} .
\end{equation}
Note that the CMN yields another formulation for the CCNR criterion:
\begin{equation}
\forall \rho \in \mathcal{S}, \quad \mathcal{M}_{h=1,p=1} \leq 1 ,
\end{equation}
where $ \mathcal{S} $ denotes the set of separable states. The CMN also allows for a new formulation of the CM criterion~\cite{gittsovich2008unifying}: For any separable state in FNF, $ \mathcal{M}_{h=1,p=1} \leq \frac{1+ \sqrt{ \left( D-1 \right) \left( d-1 \right)}}{\sqrt{Dd}} $. Note the RHS is strictly smaller than $1$ iff $D \neq d$.

\section{The operator-Schmidt decomposition of a separable state}\label{operator_Schmidt_separable}
Assume $ D = d_A \geq d_B = d $. We wish to find the Schmidt coefficients of the following density matrix:
\begin{equation}\label{rho_with_Pk_Qk}
\rho = \sum_{k=1}^{n} p_k O_k \otimes Q_k .
\end{equation}

\subsection{Aside: $n=d^2$}
First, let us prove we can always assume that $n=d^2$ (however, $O_k, Q_k$ are not necessarily pure):
Suppose $ n>d^2 $. It suffices to show we can always transform \eqref{rho_with_Pk_Qk} to a similar state with $n-1$. Since the $ Q_k $ all belong to the space of $d \times d$ Hermitian matrices, they must be linearly dependent; i.e., thus, one of them (w.l.g. it is $Q_n$) may be written as a linear combination of the others:
\begin{equation}
\exists c_1, \ldots, c_{n-1} : Q_n = \sum_{k=1}^{n-1} c_k Q_k
\end{equation}
where $ \tr{Q_n} = 1 $ implies $ \sum c_k = 1 $.
Plugging this into \eqref{rho_with_Pk_Qk} yields:
\begin{align}
\rho & = \sum_{k=1}^{n-1} p_k O_k \otimes Q_k + p_n O_n \otimes \sum_{k=1}^{n-1} c_k Q_k = \nonumber\\ 
& = \sum_{k=1}^{n-1} \left( p_k O_k + p_n c_k O_n \right) \otimes Q_k = \nonumber\\
& = \sum_{k=1}^{n-1} \underbrace{ \left( p_k + p_n c_k \right) }_{\tilde{p}_k} \underbrace{ \frac{ p_k O_k + p_n c_k O_n }{p_k + p_n c_k} }_{\tilde{O}_k} \otimes Q_k .
\end{align}
To conclude the proof, one should verify $ \sum_{k=1}^{n-1} \tilde{p}_k = 1 $ and $ \tr{\tilde{O}_k}=1 $. This is straightforward so we do not show it here.

\subsection{Realignment and correlation in Bloch vector representation}
Let us write the realigned density matrix:
\begin{equation}
\rho_R = \sum_{k=1}^{n} p_k \vect{O_k} \vect{Q_k}^\dagger .
\end{equation}
Now, we shall write down $ \rho_R^\dagger \rho_R $ as a ``superoperator'' $ \hat{\mathscr{P}} $ - i.e., its operates on $ d \times d $ Hermitian operators:
\begin{equation}
\hat{\mathscr{P}} = \sum_{k=1}^{n} p_k O_k \otimes Q_k
\end{equation}
here the tensor product sign $ \otimes $ has a meaning closer to its original one, rather than its regular abuse in quantum information theory; that is, it ``wants'' to act on a $ d \times d $ Hermitian operator with the Hilbert-Schmidt inner product as follows:
\begin{equation}
\left( A \otimes B \right) C = \braket{B,C} A = \tr{B^\dagger C} A
\end{equation}
where $ B,C $ are both all $ d \times d $ (Hermitian) operators.
For the sake of simplicity, we switch to the Bloch representation of the operators, satisfying the following properties:
\begin{enumerate}
	\item Each operator $Q_k$ is written as $ Q_k = \frac{1}{\sqrt{d}} q_k^\mu \hat{\sigma}_\mu $, where $ \mu =0, 1, \ldots, d^2-1 $. $ \hat{\sigma}_0 = \mathbb{1}_d / \sqrt{d} $, and the other $ \hat{\sigma}_i $ are (traceless) $ d \times d $ Hermitian operators s.t. all $ \hat{\sigma}_\mu $ are an orthogonal set (w.r.t. the Hilbert-Schmidt inner product), satisfying: 
	\begin{equation*}
	\tr{ \hat{\sigma}_\nu \hat{\sigma}_\mu } = \delta_{\nu \mu}
	\end{equation*}
	and the $ q_k^\mu $ are real numbers, given by:
	\begin{equation}
	q_k^\mu = \sqrt{d} \cdot \tr{\hat{\sigma}_\mu Q_k} .
	\end{equation}
	Note that $q_k^0 = \tr{Q_k} = 1$.
	\item This notation allows one to compute the Hilbert-Schmidt inner product of two operators $ U = \frac{1}{\sqrt{d}} u^\nu \hat{\sigma}_\nu $ and $ V = \frac{1}{\sqrt{d}} v^\mu \hat{\sigma}_\mu $ as follows:
	\begin{align}
	\braket{U,V}_{HS} = & \tr{UV} = \frac{1}{d} \tr{ u^\nu \hat{\sigma}_\nu v^\mu \hat{\sigma}_\mu } = \nonumber\\
	& = \frac{1 }{d} u^\nu v^\mu \tr{ \hat{\sigma}_\nu \hat{\sigma}_\mu } = \frac{1}{d} u^\mu v_\mu . \nonumber
	\end{align}
\end{enumerate}
And similarly for the operators $O_k$ (where $d$ is replaced with $D \defeq \max \left\{ d_A, d_B \right\} = d_A $):
\begin{equation}
O_k = \frac{1}{\sqrt{D}} o_k^\gamma \hat{\xi}_\gamma , \quad \tr{\hat{\xi}_\gamma \hat{\xi}_\eta} = \delta_{\gamma \eta} , \quad o_k^\gamma = \sqrt{D} \cdot \tr{\hat{\xi}_\gamma O_k} .
\end{equation}
Once Hermitian operators are represented by column vectors (using the bases $ \left\{ \hat{\xi}_\gamma \right\}, \left\{ \hat{\sigma}_\mu \right\} $), the superoperator $ \hat{\mathscr{P}} $ may once again be written as a $ D^2 \times d^2 $ matrix:
\begin{equation}
\mathcal{C}^{\gamma \mu} = \frac{1}{\sqrt{Dd}} \sum_{k=1}^{n} p_k o_k^\gamma q_k^\mu = \frac{1}{\sqrt{Dd}} \mathcal{O P Q}^T,
\end{equation}
where $\mathcal{O}^{\gamma}_{\;\; l} \defeq o_l^\gamma , \mathcal{Q}^{\mu}_{\;\;k} \defeq q_k^\mu $ are the matrices with columns comprised of the Bloch vectors of $ O_l, Q_k $ respectively; and $ \mathcal{P} \defeq \diag \left[ p_1, \ldots, p_n \right] $. Later on, it shall be useful to consider the matrix $ \mathcal{R} \defeq \mathcal{C}^T \mathcal{C} $, since its eigenvalues are the squared singular values of $ \mathcal{C} $:
\begin{equation}
\mathcal{R} = \frac{1}{D d} \left( \mathcal{O P Q}^T \right)^T \mathcal{O P Q}^T = \frac{1}{ D d } \mathcal{Q P O}^T \mathcal{O P Q}^T .
\end{equation}

\subsection{Separability}
Up until this point we still haven't used the \textit{separability} of $\rho$; it manifests in the fact that the operators $O_k, Q_l$ all represent states, implying:
\begin{equation}\label{bloch_constraints}
\forall k, \begin{cases}
\tr{O_k} = \tr{Q_k} = 1 \quad \Rightarrow \quad o_k^0 = q_k^0 = 1 \\
\tr{O_k^2}\leq 1 \land \tr{Q_k^2} \leq 1 \, \Rightarrow \, o_k^\mu o_k^\mu \leq D, \, q_k^\nu q_k^\nu \leq d ,
\end{cases}
\end{equation}
where only the Greek indices are summed upon. I.e., the first row of $ \mathcal{O,Q} $ is all ones; and the main diagonals of $ \frac{1}{D} \mathcal{O}^T \mathcal{O} $, $ \frac{1}{d} \mathcal{Q}^T \mathcal{Q} $ are bounded by one. Let us denote $ \mathcal{O_+,Q_+} $ as the matrices obtained by removing the all-ones first rows from $ \mathcal{O,Q} $ respectively. It would be useful to unify the two conditions, by writing down the summation explicitly:
\begin{align}
\forall k, \quad 1 & \geq \tr{O_k^2} = \frac{1}{D} \sum_{\mu=0}^{D^2-1} o_k^\mu o_k^\mu = \nonumber\\
& = \frac{1}{D} \left( 1+ \sum_{\mu=1}^{D^2-1} o_k^\mu o_k^\mu \right) = \frac{1}{D} \left( 1+ \left[ \mathcal{O}_+^T \mathcal{O}_+ \right]_{kk} \right) ,
\end{align}
implying:
\begin{equation}
\forall k, \quad \left[ \mathcal{O}_+^T \mathcal{O}_+ \right]_{kk} \leq D-1
\end{equation}
and similarly:
\begin{equation}
\forall k, \quad \left[ \mathcal{Q}_+^T \mathcal{Q}_+ \right]_{kk} \leq d-1 .
\end{equation}

Finally, we note that:
\begin{equation}
r^\gamma = \mathcal{C}^{\gamma 0} = \frac{1}{\sqrt{Dd}} \sum_k p_k o_k^\gamma
\end{equation}
implying:
\begin{equation}
\bm{r} = \frac{1}{\sqrt{Dd}} \mathcal{O}_+ \bm{p} .
\end{equation}
This is unsurprising, since $ \left[1,\bm{r}^T \right] $ is the Bloch vector of $ \rho_A = \sum_k p_k O_k $. Similarly:
\begin{equation}
\bm{s} = \frac{1}{\sqrt{Dd}} \mathcal{Q}_+ \bm{p} .
\end{equation}

\subsection{FNF}
Let us assume that Alice and Bob choose their orthonormal observables such that $ A_1 = \frac{1}{\sqrt{d_A}} \mathbb{1}_{d_A} $ and $ B_1 = \frac{1}{\sqrt{d_B}} \mathbb{1}_{d_B} $, i.e. the trivial measurements. Note this implies that all the other observables $ A_i, B_j $ are traceless. Given this assumption, we are motivated to introduce the following notation (similar to~\cite{vicente2007separability}): 
\begin{equation}
\mathcal{C} = \begin{bmatrix}
1/\sqrt{D d} & \bm{s}^T \\
\bm{r} & \mathcal{T}
\end{bmatrix} ,
\end{equation}
i.e.: $ r_i \defeq \braket{ A_i \otimes \mathbb{1} } / \sqrt{d_B}, s_j \defeq \braket{\mathbb{1} \otimes B_j } / \sqrt{d_A} $, and $ \mathcal{T} $ is the correlation matrix of only traceless observables. A state $\rho$ is said to be in FNF if $ \bm{r} = \bm{0} $ and $ \bm{s} = \bm{0} $. Any state may be transformed to FNF (using SLOCC), such that the original state is separable iff the transformed state is separable.

We note:
\begin{equation}\label{T_as_product}
\mathcal{T} = \frac{1}{\sqrt{d D} } \mathcal{O_+ P Q_+}^T .
\end{equation}
A recent paper~\cite{li2018necessary} has used similar ideas to construct a necessary and sufficient separability criterion; in fact, they state that a correlation matrix $ \mathcal{T} $ describes a separable state in FNF iff it admits a decomposition of the form \eqref{T_as_product}, where $ P = \diag \bm{p} $ is diagonal, real, non-negative and has unit trace; and $ \mathcal{O_+} \bm{p} = \bm{0} $, $ \mathcal{Q_+} \bm{p} = \bm{0} $. The latter conditions are related to the FNF: from \eqref{rho_with_Pk_Qk}, we observe that $\rho$ is in FNF iff $ \sum_k p_k O_k, \sum_k p_k Q_k $ are both proportional to the identity; considering this statement in Bloch vector terms readily implies these conditions.

\section{Proving the upper bounds}\label{app_proofs}
This section uses the notation and results detailed in Section \ref{operator_Schmidt_separable} of this supplemental material to prove the main results of our work.
\subsection{Preliminaries}
Observe the following:
\begin{align}\label{Holder}
& \sum_{j=1}^{d^2-1} \sigma_j \left( \mathcal{T} \right) = \frac{1}{\sqrt{Dd}} \sum_{k=1}^{d^2-1} \sigma_k \left( \mathcal{O_+ P Q_+}^T \right) = \nonumber\\
& = \frac{1}{\sqrt{Dd}} \norm{ \mathcal{O_+ \sqrt{P} } \mathcal{ \sqrt{P} Q_+^T } }_1 \leq \frac{1}{\sqrt{Dd}} \norm{ \mathcal{O_+ \sqrt{P} } }_2 \norm{ \mathcal{Q_+ \sqrt{P} } }_2
\end{align}
where the final transition follows from H\"{o}lder's inequality. Let us find bounds on the $2$-norms:
\begin{align}
\norm{ \mathcal{O_+ \sqrt{P} } }_2^2 & = \tr{ \mathcal{ \sqrt{P} O_+^T O_+ \sqrt{P} } } = \tr{  \mathcal{ P O_+^T O_+} } = \nonumber\\
& = \sum_{k=1}^{d^2} p_k \left[ \mathcal{O_+^T O_+} \right]_{kk} \leq D-1
\end{align}
and similarly,
\begin{equation}
\norm{ \mathcal{Q_+ \sqrt{P} } }_2^2 \leq d-1 .
\end{equation}
Substitution of the latter two in \eqref{Holder} yields
\begin{equation}\label{dV}
\sum_{k=1}^{d^2-1} \sigma_k \leq \sqrt{ \frac{D-1}{D} \frac{d-1}{d} } .
\end{equation}
Note this inequality is equivalent to the separability criterion defined by de Vicente in~\cite{vicente2007separability} (the dV criterion). de Vicente defines a matrix $ T $ similar to our $ \mathcal{T} $; in fact, $ T = \frac{D d}{2} \mathcal{T} $. The dV criterion states that for any separable state,
\begin{equation}\label{dV_original}
\norm{T}_{KF} \leq \sqrt{ \frac{D d \left( D-1 \right) \left(d-1 \right) }{4} } ,
\end{equation}
where $ \norm{\cdot}_{KF} $ denotes the Ky-Fan norm, i.e. the Schatten $1$-norm (also known as the trace norm or nuclear norm). \eqref{dV_original} implies:
\begin{equation}
\sum_{j=1}^{d^2-1} \sigma_j \left( \mathcal{T} \right) = \norm{\mathcal{T}}_{KF} = \frac{2}{D d} \norm{T}_{KF} \leq \sqrt{ \frac{D-1}{D} \frac{d-1}{d} } .
\end{equation}

\subsection{Proof of the upper bound of $\mathcal{ M}_{h,p=1}$}\label{app_proof_thm_1}
In this subsection, we wish to use the results of the previous section to find bounds on $ \mathcal{M}_{h, p=1} = \norm{ C_h \left( \mathcal{C}  \right) }_1 $ for separable states. Clearly, the tight upper bound for $ h = 1$ is $ \mathcal{M}_{h=1, p=1} \leq 1 $ (CCNR). Here we add three other assumptions: one regarding the domain of $h$ - $ h>1 $; another is $ D \leq d^3 $; and finally, we assume the state is in FNF.

Thus, $\sigma_{0} = 1 / \sqrt{Dd}$, and we obtain:
\begin{align}\label{bounding_p_1}
 & \mathcal{M}_{h,p=1} = S_h \left( 1 / \sqrt{Dd}, \sigma_1, \ldots, \sigma_{d^2-1} \right) = \nonumber\\
 & = \frac{1}{\sqrt{Dd}} S_{h-1} \left( \sigma_1, \ldots, \sigma_{d^2-1} \right) + S_h \left( \sigma_1, \ldots, \sigma_{d^2-1} \right)
\end{align}
Let us denote $ s \defeq \sum_{k=1}^{d^2-1} \sigma_k $ and $ \beta \defeq \frac{1}{d^2-1} \sqrt{ \frac{D-1}{D } \frac{d-1}{d} } $. Clearly $ s \leq \beta \left( d^2-1 \right) $. Moreover, the vectors $ \vec{\sigma} \defeq \left( \sigma_1, \ldots, \sigma_{d^2-1} \right) $ and $ \vec{e} \defeq \frac{s}{d^2-1} \left( 1, \ldots, 1 \right) $ both sum up to $s$; thus, $ \vec{\sigma} \succeq \vec{e} $ ($\succeq$ denotes majorization). Since the symmetric polynomials $ S_h $ are Schur concave, we obtain:
\begin{equation}
S_h \left( \sigma_1, \ldots, \sigma_{d^2-1} \right) \leq S_h \left( \frac{s}{d^2-1}, \ldots, \frac{s}{d^2-1} \right) .
\end{equation}
Next, we use the fact that $ S_h $ is monotonically increasing in each of its variables, alongside the inequality $ \frac{s}{d^2-1} \leq \beta $, to obtain:
\begin{equation}
S_h \left( \sigma_1, \ldots, \sigma_{d^2-1} \right) \leq S_h \left( \beta, \ldots, \beta \right) .
\end{equation}
Substitution in \eqref{bounding_p_1} yields:
\begin{align}
\mathcal{M}_{h,p=1} & \leq \alpha S_{h-1} \left( \beta, \ldots, \beta \right) + S_h \left( \beta, \ldots, \beta \right) = \nonumber\\
& = S_h \left( \alpha, \beta, \ldots, \beta \right) 
\end{align}
where $\beta$ is always repeated $d^2-1$ times.

\subsection{Saturating the upper bound of $\mathcal{ M}_{h,p=1}$} \label{sat_p_equals_1}
In our special construction of $\rho$ from Theorem 2, $n=d^2$ and $\forall k, p_k = 1/d^2$.
The following additional assumptions follow from $ \left\{ O_k \right\}, \left\{ Q_l \right\} $ being regular, coherent, degree-$1$ quantum designs with $r = 1$ and $d^2$ elements:
\begin{align}\label{q_designs_d_squared}
& \frac{1}{D} \left[ \mathcal{O}^T \mathcal{O} \right]_{kl} = \braket{O_k, O_l} = \begin{cases}
1 ; & k=l \\
\mu_A ; & k \neq l
\end{cases} \, , \nonumber\\ 
& \frac{1}{d} \left[ \mathcal{Q}^T \mathcal{Q} \right]_{kl} = \braket{Q_k, Q_l} = \begin{cases}
1 ; & k=l \\
\mu_B ; & k \neq l
\end{cases}
\end{align}
where $ \mu_{A/B} = \frac{d^2- d_{A/B} }{d_{A/B} \left( d^2-1 \right) } $. Coherence has the following additional implication:
\begin{align}
& \sum_{k=1}^{d^2} O_k = \frac{d^2}{D} \mathbb{1}_D , \quad \sum_{k=1}^{d^2} Q_k = d \mathbb{1}_d \quad \nonumber\\
\Rightarrow & \sum_{k=1}^{d^2} o_k^\mu = \sum_{k=1}^{d^2} q_k^\mu = \begin{cases}
d^2 ; & \mu = 0 \\
0 ; & \mu \neq 0
\end{cases}
\end{align} 
in matrix notation:
\begin{equation}\label{coherence}
\mathcal{O} \bm{1} = \mathcal{Q} \bm{1} = \left[ \begin{array}{c}
d^2 \\
0 \\
\vdots \\
0
\end{array} \right], \qquad \mathcal{O}_+ \bm{1} = \mathcal{Q}_+ \bm{1} = \bm{0} .
\end{equation}
where $ \bm{1} $ is the vector whose $ d^2 $ entries all equal $1$. Furthermore, we know that such a quantum design in dimension $ d = d_B $ is in fact a SIC-POVMs; thus, $ \left\{ Q_k \right\} $ are SIC-POVMs. 

Substituting these implications allows one to obtain:
\begin{equation}
\mathcal{R}_{\mu \nu} = \frac{1}{d^5} \sum_{k=1}^{d^2} q_k^\mu q_k^\nu + \frac{\mu_A}{d^5} \sum_{k \neq l} q_k^\mu q_l^\nu .
\end{equation}
Moreover, we have:
\begin{align}
\mathcal{R}_{00} & = \frac{1}{d^5} \sum_{k,l=1}^{d^2} \braket{ O_k, O_l } \underbrace{q_k^0 q_l^0}_{=1} = \frac{1}{d^5} \braket{ \sum_{k=1}^{d^2} O_k, \sum_{l=1}^{d^2} O_l } = \nonumber\\ 
& = \frac{1}{ D^2 d} \braket{ \mathbb{1}_D, \mathbb{1}_D } = \frac{1}{d_A d_B} .
\end{align}
and for all $ \nu \neq 0 $:
\begin{align}
\mathcal{R}_{0 \nu } & = \frac{1}{d^5} \sum_{k,l=1}^{d^2} \braket{ O_k, O_l } \underbrace{q_k^0}_{=1} q_l^\nu = \frac{1}{d^5} \sum_{l=1}^{d^2} \braket{ \sum_{k=1}^{d^2} O_k, O_l } q_l^\nu = \nonumber\\ 
& = \frac{1}{D d^3} \sum_{l=1}^{d^2} \underbrace{ \braket{ \mathbb{1}_D, O_l } }_{\tr{O_l} =1 } q_l^\nu = 0.
\end{align}
Similarly, for all $\mu \neq 0$, $ \mathcal{R}_{\mu 0} = 0 $. Thus, $ \lambda_0 = \mathcal{R}_{00} = \frac{1}{d_A d_B} $ is an eigenvalue. To conclude the proof, we just need to show that the submatrix of $ \mathcal{R} $ without the first row and column - i.e., $ \mathcal{T}^T\mathcal{T} $ - is the scalar matrix $\beta^2 \mathbb{1} $.

To do so, we note the following:
\begin{equation}\label{T_transposed_T}
\mathcal{T}^T\mathcal{T} = \frac{1}{Dd^5} \mathcal{Q}_+ \mathcal{O}_+^T \mathcal{O}_+ \mathcal{Q}_+^T
\end{equation}
where we have plugged $ \mathcal{P} = \frac{\mathbb{1}}{d^2}$ into \eqref{T_as_product}. Our first step would be computing $ \mathcal{O}_+^T \mathcal{O}_+ $. Using \eqref{q_designs_d_squared} and recalling that $ \mathcal{O}_+ $ is simply $ \mathcal{O} $ with the first row of all $1$s removed, we obtain:
\begin{equation}
\left[ \mathcal{O}_+^T \mathcal{O}_+ \right]_{kl} = \left[ \mathcal{O}^T \mathcal{O} \right]_{kl} -1 = \begin{cases}
D-1 ; & k=l \\
-\frac{D-1}{d^2-1} ; & k \neq l
\end{cases}
\end{equation}
Diagonalization of this matrix is rather straightforward; it is not difficult to obtain that it has two distinct eigenvalues:
\begin{enumerate}
	\item $ \lambda_0 = $ with multiplicity $1$, where the eigenspace is spanned by $\bm{1} \defeq \left(1, \ldots, 1 \right)^T $; and -
	\item $\lambda_1 = \frac{d^2 \left( D-1 \right) }{d^2-1} $ with multiplicity $d^2-1$ and eigenspace $ \Lambda \defeq \left( \spn \left\{ \bm{1} \right\} \right)^\perp $.
\end{enumerate}
This demonstrates that $ \mathcal{O}_+^T \mathcal{O}_+ $ behaves as a scalar matrix, when its domain is restricted to $ \Lambda \subset \mathbb{R}^{d^2} $. Thus, our next step would be showing that $ \mathcal{Q}_+^T : \mathbb{R}^{d^2-1} \rightarrow \mathbb{R}^{d^2} $ performs exactly this restriction.

In other words, we wish to prove that $ \Ima \mathcal{Q}_+^T = \Lambda $. First note $ \mathcal{Q}_+^T $ has an empty kernel, since otherwise there exists a nonzero vector orthogonal to all vectors in $ \left\{ \bm{q}_k^+ \right\} $; this would have implied that $ \left\{ \bm{q}_k^+ \right\} $ do not span the entire $d^2-1$-dimensional space of traceless Hermitian $d \times d$ operators, contradicting them comprising a SIC-POVM. Thus $ \ker \mathcal{Q}_+^T = 0$, and from the rank-nullity theorem $ \mathcal{Q}_+^T $ must have rank $d^2-1$. Hence, demonstrating that $ \Ima \mathcal{Q}_+^T \subset \Lambda $ would complete the proof. Let $\bm{v} \in \mathbb{R}^{d^2-1} $; indeed, direct computation yields:
\begin{equation}
\bm{1} \cdot \mathcal{Q}_+^T \bm{v} = \mathcal{Q}_+ \bm{1} \cdot \bm{v} = \bm{0} ,
\end{equation}
where we have used \eqref{coherence}. Thus, for all $\bm{v} \in \mathbb{R}^{d^2-1} $ we have:
\begin{equation}
\mathcal{O}_+^T \mathcal{O}_+ \mathcal{Q}_+^T \bm{v} = \frac{d^2 \left( D-1 \right) }{d^2-1} \mathcal{Q}_+^T \bm{v} ,
\end{equation}
implying,
\begin{equation}\label{almost_done}
\mathcal{O}_+^T \mathcal{O}_+ \mathcal{Q}_+^T = \frac{d^2 \left( D-1 \right) }{d^2-1} \mathcal{Q}_+^T .
\end{equation}
To complete the proof, we note that since $ \left\{ Q_k \right\} $ comprises a SIC-POVM, the columns of $ \mathcal{Q}_+ $ form a (real) equiangular tight frame in dimension $d^2-1$ with $d^2$ elements~\cite{waldron2018introduction}; thus, its \textit{frame operator} $ \mathcal{Q}_+ \mathcal{Q}_+^T $ is scalar; more specifically, it satisfies:
\begin{equation}
\mathcal{Q}_+ \mathcal{Q}_+^T = A \mathbb{1}_{d^2-1} , 
\end{equation}
where $A$ is readily found by taking the trace of both sides:
\begin{equation}\label{A}
\left( d^2-1 \right) A = \tr{ \mathcal{Q}_+ \mathcal{Q}_+^T } = \tr{ \mathcal{Q}_+^T \mathcal{Q}_+ } = d^2 \left( d-1 \right) ,
\end{equation}
where we have used \eqref{q_designs_d_squared} again. Multiplying \eqref{almost_done} by $ \mathcal{Q}_+ $ from the left yields:
\begin{align}
\mathcal{Q}_+ \mathcal{O}_+^T \mathcal{O}_+ \mathcal{Q}_+^T & = \frac{d^2 \left( D-1 \right) }{d^2-1} \mathcal{Q}_+ \mathcal{Q}_+^T = \nonumber\\ 
& = \frac{d^2 \left( D-1 \right) }{d^2-1} \frac{d^2 \left( d-1 \right) }{d^2-1} \mathbb{1}_{d^2-1} ,
\end{align}
which, when plugged into \eqref{T_transposed_T}, concludes the proof.

\subsection{Proof of the upper bound of $\mathcal{ M}_{h,p=\infty}$}\label{app_proof_thm_3}
In this subsection we prove the bound on $ \mathcal{M}_{h, p=\infty} = \norm{ C_h \left( \mathcal{C}  \right) }_\infty $ for separable states in FNF. Clearly, the tight upper bound for $ h = 1$ is $ \mathcal{M}_{h=1, p=\infty} \leq 1 $. 

Thus, $\sigma_{0} = 1 / \sqrt{Dd}$, and we obtain:
\begin{align}\label{bounding_p_infty}
& \mathcal{M}_{h,p=\infty} = \norm{ C_h \left( \mathcal{C} \right) }_\infty = \frac{1}{\sqrt{ Dd }} \norm{ C_{h-1} \left( \mathcal{T} \right) }_\infty = \nonumber\\ 
& = \frac{1}{\sqrt{ Dd }} \norm{ C_{h-1} \left( \frac{1}{\sqrt{ Dd }} \mathcal{O_+ P Q_+}^T \right) }_\infty = \nonumber\\
& = \left( Dd \right)^{-h/2} \prod_{k=1}^{h-1} \sigma_k \left( \mathcal{O_+ P Q_+}^T \right) \leq \nonumber\\
& \leq \left( Dd \right)^{-h/2} \left( h-1 \right)^{-\left( h-1 \right)} \left[ \sum_{k=1}^{h-1} \sigma_k \left( \mathcal{O_+ P Q_+}^T \right) \right]^{h-1} .
\end{align}
Let us find a bound on the sum:
\begin{align}\label{bound_on_sum}
\sum_{k=1}^{h-1} \sigma_k \left( \mathcal{O_+ P Q_+}^T \right) & \leq \sum_{k=1}^{d^2} \sigma_k \left( \mathcal{O_+ P Q_+}^T \right) \leq \nonumber\\ 
& \leq \sqrt{\left( D-1 \right) \left( d-1 \right)}
\end{align}
where we have used \eqref{dV}. To conclude, substituting in \eqref{bounding_p_infty} obtains the bound:
\begin{equation}
\mathcal{M}_{h,p=\infty} \leq \frac{1}{\sqrt{Dd}} \left[\frac{D-1}{D \left(h-1\right) } \frac{d-1}{d \left(h-1\right) } \right]^{\frac{h-1}{2} } .
\end{equation}

\subsection{Evidence to support Theorem 3 without assuming FNF}\label{app_evidence}
$ \mathcal{M}_{h,p=\infty} $ is a monotonically non-decreasing differentiable function of the singular values $ \vec{\sigma} = \left( \sigma_0, \ldots, \sigma_{d^2-1} \right)$. The constraints on the domain of $\vec{\sigma}$ are rather complicated and we do not know them all. However, we know \textit{some} of them:
\begin{align}
& \sigma_0 \geq 1/\sqrt{Dd} \label{constr_0}\\
\forall j \in \left\{ 1, \ldots, d^2-1 \right\}, \quad & \sigma_j \geq 0 \label{constr_1}\\
& \sum_{j=0}^{d^2} \sigma_j \leq 1 \label{constr_2}\\ 
& \sum_{j=1}^{d^2-1} \sigma_j \leq \frac{D-1}{D} \frac{d-1}{d} .\label{constr_3}
\end{align}
Furthermore, we know from numerical simulations that \eqref{constr_2} and \eqref{constr_3} cannot be saturated simultaneously (for $Dd \leq h^2 $); in fact, it seems that if the latter is saturated, then the state must be in FNF (thus saturating \eqref{constr_0} instead). \textit{Assume} that this statement holds in general, and that no other constraints on $ \vec{\sigma} $ are relevant for global maxima analysis of $ \mathcal{M}_{h,p=\infty} $ - i.e., no other constraints need be saturated to obtain its global maxima; then, the theorem holds.

Since $ \mathcal{M}_{h,p=\infty} $ is monotonically increasing, one of the constraints \eqref{constr_2},\eqref{constr_3} must be saturated in a global maximum; otherwise, any one of the $ \sigma_k $ could be increased, thus increasing the value of $ \mathcal{M}_{h,p=\infty} $ without leaving the domain. According to our assumption, if \eqref{constr_3} is saturated the state is in FNF, which is the case we already treated. Thus, assume \eqref{constr_2} is saturated. If more than $h$ singular values are nonzero, the point cannot be a global maximum, since we can increase the largest singular value while decreasing the smallest nonzero singular value, thus leaving \eqref{constr_2} saturated while increasing $ \mathcal{M}_{h,p=\infty} $. Thus, we may treat $ \mathcal{M}_{h,p=\infty} $ as a function depending only on the $h$ largest singular values: 
\begin{equation}\label{M_h_p_func_h}
f \left( \vec{\sigma} \right) = \prod_{k=0}^{h-1} \sigma_k ,
\end{equation}
And we are currently considering a global maximum $ \vec{\sigma}' = \left( \sigma_{0}, \ldots, \sigma_{h-1} \right) $ s.t. $ \sum_{k=0}^{h-1} \sigma_k = 1 $. Clearly, non of the $ \sigma_k $ can be zero - otherwise $ \vec{\sigma}' $ is a minimum rather than a maximum. Thus, of all the above constraints, $ \vec{\sigma}' $ saturates only \eqref{constr_2}. Thus, it should be a \textit{local} maximum of the following function constructed using a Lagrange multiplier:
\begin{equation}
g \left( \vec{\sigma}, \lambda \right) \defeq \prod_{k=0}^{h-1} \sigma_k -\lambda \left( \sum_{k=0}^{h-1} \sigma_k -1 \right) .
\end{equation}
Thus, the partial derivatives with respect to $ \sigma_k $ should vanish:
\begin{equation}
0 = \frac{\partial g}{\partial \sigma_l} = \prod_{k \neq l} \sigma_k -\lambda
\end{equation}
implying that for all $l$, $ \prod_{k \neq l} \sigma_k = \lambda $; but that could only happen if $ \sigma_0 = \ldots = \sigma_{h-1} = 1 / h $. Substituting $ \sigma_0 = 1/h $ in \eqref{constr_0} would have implied $ h \leq \sqrt{Dd} $. If this is an equality, we are again in FNF; otherwise, it contradicts one of our initial assumptions. Thus, the only possible global maximum is the one obtained in FNF, for which Theorem 3 holds.

\subsection{Saturating the upper bound of $\mathcal{M}_{h,p=\infty}$}\label{app_proof_thm_4}
In our special construction of $\rho$ from Theorem 4, $n=h$ and $\forall k, p_k = 1/h$.
Since $ \left\{ O_k \right\}, \left\{ Q_l \right\} $ are regular, coherent, degree-$1$ quantum designs with $r = 1$ and $h$ elements, \eqref{q_designs_d_squared} still holds; the only difference is that in this case, 
$ \mu_{A/B} = \frac{h- d_{A/B} }{d_{A/B} \left( h-1 \right) } $. As before, coherence has an additional implication:
\begin{align}
& \sum_{k=1}^{h} O_k = \frac{h}{D} \mathbb{1}_D , \quad \sum_{k=1}^{h} Q_k = \frac{h}{d} \mathbb{1}_d \nonumber\\ 
\Rightarrow \quad & \sum_{k=1}^{h} o_k^\mu = \sum_{k=1}^{h} q_k^\mu = \begin{cases}
h ; & \mu = 0 \\
0 ; & \mu \neq 0
\end{cases}
\end{align} 
in matrix notation:
\begin{equation}\label{coherence_h}
\mathcal{O} \bm{1} = \mathcal{Q} \bm{1} = \left[ \begin{array}{c}
h \\
0 \\
\vdots \\
0
\end{array} \right], \qquad \mathcal{O}_+ \bm{1} = \mathcal{Q}_+ \bm{1} = \bm{0} ,
\end{equation}
where $ \bm{1} $ is the vector whose $ h $ entries all equal $1$.

Substituting these implications allows one to obtain:
\begin{equation}
\mathcal{R}_{\mu \nu} = \frac{1}{h^2 d} \sum_{k=1}^{h} q_k^\mu q_k^\nu + \frac{\mu_A}{h^2 d} \sum_{k \neq l} q_k^\mu q_l^\nu .
\end{equation}
Moreover, we have:
\begin{align}
\mathcal{R}_{00} & = \frac{1}{h^2 d} \sum_{k,l=1}^{h} \braket{ O_k, O_l } \underbrace{q_k^0 q_l^0}_{=1} = \frac{1}{h^2 d} \braket{ \sum_{k=1}^{h} O_k, \sum_{l=1}^{h} O_l } = \nonumber\\ 
& = \frac{1}{ D^2 d} \braket{ \mathbb{1}_D, \mathbb{1}_D } = \frac{1}{d_A d_B} .
\end{align}
and for all $ \nu \neq 0 $:
\begin{align}
& \mathcal{R}_{0 \nu } = \frac{1}{h^2 d} \sum_{k,l=1}^{h} \braket{ O_k, O_l } \underbrace{q_k^0}_{=1} q_l^\nu = \nonumber\\
& = \frac{1}{h^2 d} \sum_{l=1}^{h} \braket{ \sum_{k=1}^{h} O_k, O_l } q_l^\nu = \frac{1}{h D d} \sum_{l=1}^{h} \underbrace{ \braket{ \mathbb{1}_D, O_l } }_{\tr{O_l} =1 } q_l^\nu = 0.
\end{align}
Similarly, for all $\mu \neq 0$, $ \mathcal{R}_{\mu 0} = 0 $. Thus, $ \lambda_0 = \mathcal{R}_{00} = \frac{1}{d_A d_B} $ is an eigenvalue. To conclude the proof, we need to show that the submatrix of $ \mathcal{R} $ without the first row and column - again, $ \mathcal{T}^T \mathcal{T}$ - is composed of two diagonal blocks, one being a nontrivial scalar matrix and the other is the zero matrix. 

We commence in a manner similar to what we have done in subsection \ref{sat_p_equals_1} - writing down $ \mathcal{T}^T \mathcal{T}$:
\begin{equation}
\mathcal{T}^T \mathcal{T} = \frac{1}{D d h^2} \mathcal{Q}_+ \mathcal{O}_+^T \mathcal{O}_+ \mathcal{Q}_+^T ,
\end{equation}
and computing $ \mathcal{O}_+^T \mathcal{O}_+ $:
\begin{equation}
\left[ \mathcal{O}_+^T \mathcal{O}_+ \right]_{kl} = \left[ \mathcal{O}^T \mathcal{O} \right]_{kl} -1 = \begin{cases}
D-1 ; & k = l \\
-\frac{D-1}{h-1} ; & k \neq l
\end{cases}
\end{equation}
This $ h \times h $ matrix has the eigenvalues:
\begin{enumerate}
	\item $ \lambda_0 = $ with multiplicity $1$, where the eigenspace is spanned by $\bm{1} \defeq \left(1, \ldots, 1 \right)^T $; and -
	\item $\lambda_1 = \frac{h \left( D-1 \right) }{h-1} $ with multiplicity $h-1$ and eigenspace $ \Lambda \defeq \left( \spn \left\{ \bm{1} \right\} \right)^\perp $.
\end{enumerate}
Next, we consider $ \mathcal{Q}_+^T : \mathbb{R}^{d^2-1} \rightarrow \mathbb{R}^h $. This time it does not have an empty kernel. However, it turns out we need not show that $ \Ima \mathcal{Q}_+^T = \Lambda $. It suffices to show $ \Ima \mathcal{Q}_+^T \subset \Lambda $. Indeed, this follows simply as before:
\begin{equation}
\forall \bm{v} \in \mathbb{R}^{d^2-1}, \quad \bm{1} \cdot \mathcal{Q}_+^T \bm{v} = \mathcal{Q}_+ \bm{1} \cdot \bm{v} = \bm{0} ,
\end{equation}
where the last transition is just \eqref{coherence_h}. Thus we obtain:
\begin{equation}
\mathcal{O}_+^T \mathcal{O}_+ \mathcal{Q}_+^T = \frac{h \left( D-1 \right) }{h-1} \mathcal{Q}_+^T .
\end{equation}
To conclude the proof, we must analyze $ \mathcal{Q}_+ \mathcal{Q}_+^T $. Since for $ h < d^2 $ the projections $ \left\{ Q_k \right\} $ do not comprise a SIC-POVM, $ \mathcal{Q}_+ \mathcal{Q}_+^T $ is no longer a frame operator of a tight frame, and thus not necessarily scalar. However, we may use the fact that $ \mathcal{Q}_+ \mathcal{Q}_+^T $ and $ \mathcal{Q}_+^T \mathcal{Q}_+ $ have the same \textit{nonzero} eigenvalues (i.e., squares of the singular values of $ \mathcal{Q}_+ $). Therefore, our next step would be computing $ \mathcal{Q}_+^T \mathcal{Q}_+ $:
\begin{equation}
\left[ \mathcal{Q}_+^T \mathcal{Q}_+ \right]_{kl} = \left[ \mathcal{Q}^T \mathcal{Q} \right]_{kl} -1 = \begin{cases}
d-1 ; & k = l \\
-\frac{d-1}{h-1} ; & k \neq l
\end{cases}
\end{equation}
As before, it is straightforward to note this $ h \times h $ matrix has two eigenvalues: $ \lambda_0 = 0 $ with multiplicity $1$, and $ \lambda_1 = \frac{h \left( d-1 \right)}{ h-1 } $ with multiplicity $h-1$. Thus, $ \mathcal{Q}_+ \mathcal{Q}_+^T $ has the eigenvalues $ \lambda_1 $ with multiplicity $h-1$, and $0$ with multiplicity $ d^2 -h $.

To conclude, we observed the following:
\begin{enumerate}
	\item $ \mathcal{T}^T \mathcal{T} = \frac{1}{D d h^2} \frac{h \left( D-1 \right) }{h-1} \mathcal{Q}_+ \mathcal{Q}_+^T $,
	\item $ \mathcal{Q}_+ \mathcal{Q}_+^T $ has precisely $h-1$ nonzero eigenvalues, which all equal $ \lambda_1 = \frac{h \left( d-1 \right)}{ h-1 } $.
\end{enumerate}
Thus, $ \mathcal{T}^T \mathcal{T} $ also has $h-1$ nonzero eigenvalues, which all equal $ \lambda' = \frac{D-1}{ D \left( h-1 \right)} \frac{d-1}{ d \left( h-1 \right)} $. Consequentially, the $h$ largest singular values of $ \mathcal{C} $ are $ \sigma_{0} = 1 / \sqrt{D d} $ with multiplicity $1$, and $\sqrt{\lambda'} $ with multiplicity $h-1$; and the CMN is their product, that is:
\begin{equation}
\mathcal{M}_{h,p=\infty} = \frac{1}{\sqrt{Dd} } \left[ \frac{D-1}{ D \left( h-1 \right)} \frac{d-1}{ d \left( h-1 \right)} \right]^{\frac{h-1}{2}}
\end{equation}
which concludes the proof.

\section{Relation to Quantum Discord}\label{app_discord}
It is known \cite{bera2017quantum} that for any given state $\rho$, the quantum discord $ \mathcal{D}^A $ is zero if and only if there exists a local measurement on $A$ that does not disturb the state. Here, a measurement corresponds to any orthonormal basis $ \left\{ \Pi_l \right\} $ of $ \mathcal{H}_A $, where $ \Pi_l = \ket{l} \bra{l} $ and $ \braket{ k \vert l } = \delta_{kl} $. Measuring the state $ \rho $ in this basis transforms it by:
\begin{equation}
    \rho \rightarrow \rho' = \sum_{l=1}^{d_A} \left( \Pi_l \otimes \mathbb{1} \right) \rho \left( \Pi_l \otimes \mathbb{1} \right) .
\end{equation}
Let us find the transformation ondergone by the correlation matrix:
\begin{align}\label{trans_corr}
    & \mathcal{C'}_{ij} = \tr{\rho' A_i \otimes B_j} = \nonumber\\
    & = \sum_{l=1}^{d_A} \tr{ \left( \Pi_l \otimes \mathbb{1} \right) \rho \left( \Pi_l \otimes \mathbb{1} \right) A_i \otimes B_j } = \nonumber\\
    & = \sum_{l=1}^{d_A} \tr{ \rho \left( \Pi_l A_i  \Pi_l \right) \otimes B_j } = \sum_{l=1}^{d_A} \braket{l \vert A_i \vert l} \tr{ \rho \Pi_l \otimes B_j } .
\end{align}
Since $ \left\{ A_i \right\}$ comprise an orthonormal basis of the space of Hermitian matrices over $ \mathcal{H}_A $, we may write:
\begin{equation}
    \Pi_l = \sum_{k=1}^{d_A^2} \tr{A_k \Pi_l} A_k =  \sum_{k=1}^{d_A^2} \braket{l \vert A_k \vert l} A_k .
\end{equation}
Plugging into \eqref{trans_corr}, we obtain:
\begin{align}
    & \mathcal{C'}_{ij} = \sum_{k,l=1}^{d_A^2} \braket{l \vert A_i \vert l} \braket{l \vert A_k \vert l} \underbrace{ \tr{ \rho A_k \otimes B_j } }_{\mathcal{C}_{ij} } .
\end{align}
Hence, the post-measurement correlation matrix is given by $ \mathcal{C'} = \mathcal{AC} $, where $ \mathcal{A} $ is a $ d_A^2 \times d_A^2 $ real matrix given by:
\begin{equation}
    \mathcal{A}_{ik} \defeq \sum_{l=1}^{d_A} \braket{l \vert A_i \vert l} \braket{l \vert A_k \vert l} .
\end{equation}
Equivalently, $ \mathcal{A} $ may be written as $ X X^T $, where:
\begin{equation}
    X_{il} = \braket{l \vert A_i \vert l} , \quad 1 \leq i \leq d_A^2 , \, 1 \leq l \leq d_A .
\end{equation}
Note that $ \rank X = d_A$, since its columns are the components of the orthonormal set $ \ket{l} \bra{l} $ in the orthonormal basis $ \left\{ A_i \right\} $ (w.r.t. the Hilbert-Schmidt inner product). In fact, this logic also shows that the columns of $X$ form an orthonormal basis. Hence $ X^T X = \mathbb{1}_{d_A} $, and we may observe that
\begin{equation}
    \mathcal{A}^2 = X \underbrace{ X^T X }_{\mathbb{1}} X^T = \mathcal{A} ,
\end{equation}
i.e. $ \mathcal{A} $ is a rank-$d_A$ projection matrix. However, not \textit{every} $ d_A^2 \times d_A^2 $ projection matrix with rank $d_A$ is obtained by this construction from some orthonormal basis for $ \mathcal{H}_A $.

Note that the construction of $ \mathcal{A} $ is exactly the same as in Theorem 1 of \cite{luo2010geometric}. Since $ \mathcal{M}_{h=1,p=2} = \tr{C C^T} $ for any matrix $C$ (this is the sum of squared singular values), it is clear that $ \mathcal{D}_G^A = \mathcal{D}_{h=1,p=2}^A $.

Let us prove Theorem 5. First, by Theorem 6.7(7) in \cite{hiai2014introduction}, for all $k \in \left\{ 1, \ldots, d_A^2 \right\} $ we have
\begin{equation}
    \sigma_k \left( \mathcal{AC} \right) \leq \norm{ \mathcal{A} }_1 \sigma_k \left( \mathcal{C} \right) ,
\end{equation}
where $ \norm{ \mathcal{A} }_1 = \max_j \sigma_j \left( \mathcal{A} \right) = 1 $, as $ \mathcal{A} $ is a projection. Therefore, $ \sigma_k \left( \mathcal{AC} \right) \leq \sigma_k \left( \mathcal{C} \right) $ for all $k$, and we conclude that $ \mathcal{M}_{h,p} \left( \rho \right) \geq \mathcal{M}_{h,p} \left( \rho' \right) $ for all $h,p$, using the fact that the CMNs are all monotonically non-decreasing w.r.t. the singular values $ \sigma_k $.

Now, suppose $\rho$ has zero discord. As we have already noted, there must be a measurement that does not disturb the state - i.e., there exists a matrix $ \mathcal{A} $ such that $ \mathcal{AC} = \mathcal{C} $. Then, for this choice of measurement, we have $ \mathcal{M}_{h,p} \left( \rho \right) - \mathcal{M}_{h,p} \left( \rho' \right) = 0 $. By the non-decreasing property for the CMN we have proven above, this is indeed the maximum, hence $ \mathcal{D}_{h \leq 2, p} \left( \rho \right) = 0 $.

To prove the converse, suppose there exists some measurement $\Pi^A$ that changes the state but does not change the singular values.
By Theorem 1 of \cite{luo2010geometric}, this implies the discord is zero. Thus, for any positive-discord state, at least one singular value of $\mathcal{AC}$ is strictly smaller than the corresponding one in $ \mathcal{C} $. This would decrease any \textit{nonzero} monomial of degree $h$ in which it appears; and such a monomial always exists if there are at least $h$ nonzero singular values. Since we are assuming the state has discord, the rank of $\mathcal{C}$ must be at least two (this can be observed, e.g. using the condition described in \cite{dakic2010necessary}). Thus, we have proven that $ \mathcal{D}_{h \leq 2, p} \left( \rho \right) > 0 $ for any positive-discord state $\rho$.

The family of states depicted in Figure 2 is given by \cite{virzi2019optimal}:
\begin{equation}
    \rho \left( q,r \right) = \begin{bmatrix}
    0 & 0 & 0 & 0 \\
    0 & q & -r\sqrt{q \left( 1-q \right) } & 0 \\
    0 & -r\sqrt{q \left( 1-q \right) } & 1-q & 0 \\
    0 & 0 & 0 & 0
    \end{bmatrix} .
\end{equation}

\bibliographystyle{apsrev4-1}
\bibliography{ArticleReview}

\begin{thebibliography}{69}%
\makeatletter
\providecommand \@ifxundefined [1]{%
 \@ifx{#1\undefined}
}%
\providecommand \@ifnum [1]{%
 \ifnum #1\expandafter \@firstoftwo
 \else \expandafter \@secondoftwo
 \fi
}%
\providecommand \@ifx [1]{%
 \ifx #1\expandafter \@firstoftwo
 \else \expandafter \@secondoftwo
 \fi
}%
\providecommand \natexlab [1]{#1}%
\providecommand \enquote  [1]{``#1''}%
\providecommand \bibnamefont  [1]{#1}%
\providecommand \bibfnamefont [1]{#1}%
\providecommand \citenamefont [1]{#1}%
\providecommand \href@noop [0]{\@secondoftwo}%
\providecommand \href [0]{\begingroup \@sanitize@url \@href}%
\providecommand \@href[1]{\@@startlink{#1}\@@href}%
\providecommand \@@href[1]{\endgroup#1\@@endlink}%
\providecommand \@sanitize@url [0]{\catcode `\\12\catcode `\$12\catcode
  `\&12\catcode `\#12\catcode `\^12\catcode `\_12\catcode `\%12\relax}%
\providecommand \@@startlink[1]{}%
\providecommand \@@endlink[0]{}%
\providecommand \url  [0]{\begingroup\@sanitize@url \@url }%
\providecommand \@url [1]{\endgroup\@href {#1}{\urlprefix }}%
\providecommand \urlprefix  [0]{URL }%
\providecommand \Eprint [0]{\href }%
\providecommand \doibase [0]{http://dx.doi.org/}%
\providecommand \selectlanguage [0]{\@gobble}%
\providecommand \bibinfo  [0]{\@secondoftwo}%
\providecommand \bibfield  [0]{\@secondoftwo}%
\providecommand \translation [1]{[#1]}%
\providecommand \BibitemOpen [0]{}%
\providecommand \bibitemStop [0]{}%
\providecommand \bibitemNoStop [0]{.\EOS\space}%
\providecommand \EOS [0]{\spacefactor3000\relax}%
\providecommand \BibitemShut  [1]{\csname bibitem#1\endcsname}%
\let\auto@bib@innerbib\@empty
\bibitem [{\citenamefont {Islam}\ \emph {et~al.}(2015)\citenamefont {Islam},
  \citenamefont {Ma}, \citenamefont {Preiss}, \citenamefont {Tai},
  \citenamefont {Lukin}, \citenamefont {Rispoli},\ and\ \citenamefont
  {Greiner}}]{islam2015measuring}%
  \BibitemOpen
  \bibfield  {author} {\bibinfo {author} {\bibfnamefont {R.}~\bibnamefont
  {Islam}}, \bibinfo {author} {\bibfnamefont {R.}~\bibnamefont {Ma}}, \bibinfo
  {author} {\bibfnamefont {P.~M.}\ \bibnamefont {Preiss}}, \bibinfo {author}
  {\bibfnamefont {M.~E.}\ \bibnamefont {Tai}}, \bibinfo {author} {\bibfnamefont
  {A.}~\bibnamefont {Lukin}}, \bibinfo {author} {\bibfnamefont
  {M.}~\bibnamefont {Rispoli}}, \ and\ \bibinfo {author} {\bibfnamefont
  {M.}~\bibnamefont {Greiner}},\ }\href@noop {} {\bibfield  {journal} {\bibinfo
   {journal} {Nature}\ }\textbf {\bibinfo {volume} {528}},\ \bibinfo {pages}
  {77} (\bibinfo {year} {2015})}\BibitemShut {NoStop}%
\bibitem [{\citenamefont {Amico}\ \emph {et~al.}(2008)\citenamefont {Amico},
  \citenamefont {Fazio}, \citenamefont {Osterloh},\ and\ \citenamefont
  {Vedral}}]{amico2008entanglement}%
  \BibitemOpen
  \bibfield  {author} {\bibinfo {author} {\bibfnamefont {L.}~\bibnamefont
  {Amico}}, \bibinfo {author} {\bibfnamefont {R.}~\bibnamefont {Fazio}},
  \bibinfo {author} {\bibfnamefont {A.}~\bibnamefont {Osterloh}}, \ and\
  \bibinfo {author} {\bibfnamefont {V.}~\bibnamefont {Vedral}},\ }\href@noop {}
  {\bibfield  {journal} {\bibinfo  {journal} {Rev. Mod. Phys.}\ }\textbf
  {\bibinfo {volume} {80}},\ \bibinfo {pages} {517} (\bibinfo {year}
  {2008})}\BibitemShut {NoStop}%
\bibitem [{\citenamefont {Jurcevic}\ \emph {et~al.}(2014)\citenamefont
  {Jurcevic}, \citenamefont {Lanyon}, \citenamefont {Hauke}, \citenamefont
  {Hempel}, \citenamefont {Zoller}, \citenamefont {Blatt},\ and\ \citenamefont
  {Roos}}]{jurcevic2014quasiparticle}%
  \BibitemOpen
  \bibfield  {author} {\bibinfo {author} {\bibfnamefont {P.}~\bibnamefont
  {Jurcevic}}, \bibinfo {author} {\bibfnamefont {B.~P.}\ \bibnamefont
  {Lanyon}}, \bibinfo {author} {\bibfnamefont {P.}~\bibnamefont {Hauke}},
  \bibinfo {author} {\bibfnamefont {C.}~\bibnamefont {Hempel}}, \bibinfo
  {author} {\bibfnamefont {P.}~\bibnamefont {Zoller}}, \bibinfo {author}
  {\bibfnamefont {R.}~\bibnamefont {Blatt}}, \ and\ \bibinfo {author}
  {\bibfnamefont {C.~F.}\ \bibnamefont {Roos}},\ }\href@noop {} {\bibfield
  {journal} {\bibinfo  {journal} {Nature}\ }\textbf {\bibinfo {volume} {511}},\
  \bibinfo {pages} {202} (\bibinfo {year} {2014})}\BibitemShut {NoStop}%
\bibitem [{\citenamefont {Kaufman}\ \emph {et~al.}(2016)\citenamefont
  {Kaufman}, \citenamefont {Tai}, \citenamefont {Lukin}, \citenamefont
  {Rispoli}, \citenamefont {Schittko}, \citenamefont {Preiss},\ and\
  \citenamefont {Greiner}}]{kaufman2016quantum}%
  \BibitemOpen
  \bibfield  {author} {\bibinfo {author} {\bibfnamefont {A.~M.}\ \bibnamefont
  {Kaufman}}, \bibinfo {author} {\bibfnamefont {M.~E.}\ \bibnamefont {Tai}},
  \bibinfo {author} {\bibfnamefont {A.}~\bibnamefont {Lukin}}, \bibinfo
  {author} {\bibfnamefont {M.}~\bibnamefont {Rispoli}}, \bibinfo {author}
  {\bibfnamefont {R.}~\bibnamefont {Schittko}}, \bibinfo {author}
  {\bibfnamefont {P.~M.}\ \bibnamefont {Preiss}}, \ and\ \bibinfo {author}
  {\bibfnamefont {M.}~\bibnamefont {Greiner}},\ }\href@noop {} {\bibfield
  {journal} {\bibinfo  {journal} {Science}\ }\textbf {\bibinfo {volume}
  {353}},\ \bibinfo {pages} {794} (\bibinfo {year} {2016})}\BibitemShut
  {NoStop}%
\bibitem [{\citenamefont {Braunstein}\ and\ \citenamefont
  {Van~Loock}(2005)}]{braunstein2005quantum}%
  \BibitemOpen
  \bibfield  {author} {\bibinfo {author} {\bibfnamefont {S.~L.}\ \bibnamefont
  {Braunstein}}\ and\ \bibinfo {author} {\bibfnamefont {P.}~\bibnamefont
  {Van~Loock}},\ }\href@noop {} {\bibfield  {journal} {\bibinfo  {journal}
  {Reviews of modern physics}\ }\textbf {\bibinfo {volume} {77}},\ \bibinfo
  {pages} {513} (\bibinfo {year} {2005})}\BibitemShut {NoStop}%
\bibitem [{\citenamefont {Bello}\ \emph {et~al.}(2020)\citenamefont {Bello},
  \citenamefont {Michael}, \citenamefont {Rosenbluh}, \citenamefont {Cohen},\
  and\ \citenamefont {Pe'er}}]{bello2020complex}%
  \BibitemOpen
  \bibfield  {author} {\bibinfo {author} {\bibfnamefont {L.}~\bibnamefont
  {Bello}}, \bibinfo {author} {\bibfnamefont {Y.}~\bibnamefont {Michael}},
  \bibinfo {author} {\bibfnamefont {M.}~\bibnamefont {Rosenbluh}}, \bibinfo
  {author} {\bibfnamefont {E.}~\bibnamefont {Cohen}}, \ and\ \bibinfo {author}
  {\bibfnamefont {A.}~\bibnamefont {Pe'er}},\ }\href@noop {} {\bibfield
  {journal} {\bibinfo  {journal} {arXiv preprint arXiv:2011.08099}\ } (\bibinfo
  {year} {2020})}\BibitemShut {NoStop}%
\bibitem [{\citenamefont {Berrada}\ and\ \citenamefont
  {Abdel-Khalek}(2011)}]{berrada2011entanglement}%
  \BibitemOpen
  \bibfield  {author} {\bibinfo {author} {\bibfnamefont {K.}~\bibnamefont
  {Berrada}}\ and\ \bibinfo {author} {\bibfnamefont {S.}~\bibnamefont
  {Abdel-Khalek}},\ }\href@noop {} {\bibfield  {journal} {\bibinfo  {journal}
  {Physica E Low Dimens. Syst. Nanostruct.}\ }\textbf {\bibinfo {volume}
  {44}},\ \bibinfo {pages} {628} (\bibinfo {year} {2011})}\BibitemShut
  {NoStop}%
\bibitem [{\citenamefont {Abdel-Khalek}\ \emph {et~al.}(2012)\citenamefont
  {Abdel-Khalek}, \citenamefont {Berrada},\ and\ \citenamefont
  {Ooi}}]{abdel2012beam}%
  \BibitemOpen
  \bibfield  {author} {\bibinfo {author} {\bibfnamefont {S.}~\bibnamefont
  {Abdel-Khalek}}, \bibinfo {author} {\bibfnamefont {K.}~\bibnamefont
  {Berrada}}, \ and\ \bibinfo {author} {\bibfnamefont {C.~R.}\ \bibnamefont
  {Ooi}},\ }\href@noop {} {\bibfield  {journal} {\bibinfo  {journal} {Laser
  Phys.}\ }\textbf {\bibinfo {volume} {22}},\ \bibinfo {pages} {1449} (\bibinfo
  {year} {2012})}\BibitemShut {NoStop}%
\bibitem [{\citenamefont {Chtchelkatchev}\ \emph {et~al.}(2002)\citenamefont
  {Chtchelkatchev}, \citenamefont {Blatter}, \citenamefont {Lesovik},\ and\
  \citenamefont {Martin}}]{chtchelkatchev2002bell}%
  \BibitemOpen
  \bibfield  {author} {\bibinfo {author} {\bibfnamefont {N.~M.}\ \bibnamefont
  {Chtchelkatchev}}, \bibinfo {author} {\bibfnamefont {G.}~\bibnamefont
  {Blatter}}, \bibinfo {author} {\bibfnamefont {G.~B.}\ \bibnamefont
  {Lesovik}}, \ and\ \bibinfo {author} {\bibfnamefont {T.}~\bibnamefont
  {Martin}},\ }\href@noop {} {\bibfield  {journal} {\bibinfo  {journal}
  {Physical Review B}\ }\textbf {\bibinfo {volume} {66}},\ \bibinfo {pages}
  {161320} (\bibinfo {year} {2002})}\BibitemShut {NoStop}%
\bibitem [{\citenamefont {Wie{\'s}niak}\ \emph {et~al.}(2005)\citenamefont
  {Wie{\'s}niak}, \citenamefont {Vedral},\ and\ \citenamefont
  {Brukner}}]{wiesniak2005magnetic}%
  \BibitemOpen
  \bibfield  {author} {\bibinfo {author} {\bibfnamefont {M.}~\bibnamefont
  {Wie{\'s}niak}}, \bibinfo {author} {\bibfnamefont {V.}~\bibnamefont
  {Vedral}}, \ and\ \bibinfo {author} {\bibfnamefont {{\v{C}}.}~\bibnamefont
  {Brukner}},\ }\href@noop {} {\bibfield  {journal} {\bibinfo  {journal} {New
  J. Phys.}\ }\textbf {\bibinfo {volume} {7}},\ \bibinfo {pages} {258}
  (\bibinfo {year} {2005})}\BibitemShut {NoStop}%
\bibitem [{\citenamefont {Gonz{\'a}lez-Tudela}\ and\ \citenamefont
  {Porras}(2013)}]{gonzalez2013mesoscopic}%
  \BibitemOpen
  \bibfield  {author} {\bibinfo {author} {\bibfnamefont {A.}~\bibnamefont
  {Gonz{\'a}lez-Tudela}}\ and\ \bibinfo {author} {\bibfnamefont
  {D.}~\bibnamefont {Porras}},\ }\href@noop {} {\bibfield  {journal} {\bibinfo
  {journal} {Phys. Rev. Lett.}\ }\textbf {\bibinfo {volume} {110}},\ \bibinfo
  {pages} {080502} (\bibinfo {year} {2013})}\BibitemShut {NoStop}%
\bibitem [{\citenamefont {Tichy}\ \emph {et~al.}(2011)\citenamefont {Tichy},
  \citenamefont {Mintert},\ and\ \citenamefont
  {Buchleitner}}]{tichy2011essential}%
  \BibitemOpen
  \bibfield  {author} {\bibinfo {author} {\bibfnamefont {M.~C.}\ \bibnamefont
  {Tichy}}, \bibinfo {author} {\bibfnamefont {F.}~\bibnamefont {Mintert}}, \
  and\ \bibinfo {author} {\bibfnamefont {A.}~\bibnamefont {Buchleitner}},\
  }\href@noop {} {\bibfield  {journal} {\bibinfo  {journal} {J. Phys. B}\
  }\textbf {\bibinfo {volume} {44}},\ \bibinfo {pages} {192001} (\bibinfo
  {year} {2011})}\BibitemShut {NoStop}%
\bibitem [{\citenamefont {Sackett}\ \emph {et~al.}(2000)\citenamefont
  {Sackett}, \citenamefont {Kielpinski}, \citenamefont {King}, \citenamefont
  {Langer}, \citenamefont {Meyer}, \citenamefont {Myatt}, \citenamefont {Rowe},
  \citenamefont {Turchette}, \citenamefont {Itano}, \citenamefont {Wineland}
  \emph {et~al.}}]{sackett2000experimental}%
  \BibitemOpen
  \bibfield  {author} {\bibinfo {author} {\bibfnamefont {C.~A.}\ \bibnamefont
  {Sackett}}, \bibinfo {author} {\bibfnamefont {D.}~\bibnamefont {Kielpinski}},
  \bibinfo {author} {\bibfnamefont {B.~E.}\ \bibnamefont {King}}, \bibinfo
  {author} {\bibfnamefont {C.}~\bibnamefont {Langer}}, \bibinfo {author}
  {\bibfnamefont {V.}~\bibnamefont {Meyer}}, \bibinfo {author} {\bibfnamefont
  {C.~J.}\ \bibnamefont {Myatt}}, \bibinfo {author} {\bibfnamefont
  {M.}~\bibnamefont {Rowe}}, \bibinfo {author} {\bibfnamefont {Q.}~\bibnamefont
  {Turchette}}, \bibinfo {author} {\bibfnamefont {W.~M.}\ \bibnamefont
  {Itano}}, \bibinfo {author} {\bibfnamefont {D.~J.}\ \bibnamefont {Wineland}},
   \emph {et~al.},\ }\href@noop {} {\bibfield  {journal} {\bibinfo  {journal}
  {Nature}\ }\textbf {\bibinfo {volume} {404}},\ \bibinfo {pages} {256}
  (\bibinfo {year} {2000})}\BibitemShut {NoStop}%
\bibitem [{\citenamefont {Jaksch}\ \emph {et~al.}(1999)\citenamefont {Jaksch},
  \citenamefont {Briegel}, \citenamefont {Cirac}, \citenamefont {Gardiner},\
  and\ \citenamefont {Zoller}}]{jaksch1999entanglement}%
  \BibitemOpen
  \bibfield  {author} {\bibinfo {author} {\bibfnamefont {D.}~\bibnamefont
  {Jaksch}}, \bibinfo {author} {\bibfnamefont {H.-J.}\ \bibnamefont {Briegel}},
  \bibinfo {author} {\bibfnamefont {J.}~\bibnamefont {Cirac}}, \bibinfo
  {author} {\bibfnamefont {C.}~\bibnamefont {Gardiner}}, \ and\ \bibinfo
  {author} {\bibfnamefont {P.}~\bibnamefont {Zoller}},\ }\href@noop {}
  {\bibfield  {journal} {\bibinfo  {journal} {Phys. Rev. Lett.}\ }\textbf
  {\bibinfo {volume} {82}},\ \bibinfo {pages} {1975} (\bibinfo {year}
  {1999})}\BibitemShut {NoStop}%
\bibitem [{\citenamefont {Y{\"o}na{\c{c}}}\ \emph {et~al.}(2006)\citenamefont
  {Y{\"o}na{\c{c}}}, \citenamefont {Yu},\ and\ \citenamefont
  {Eberly}}]{yonacc2006sudden}%
  \BibitemOpen
  \bibfield  {author} {\bibinfo {author} {\bibfnamefont {M.}~\bibnamefont
  {Y{\"o}na{\c{c}}}}, \bibinfo {author} {\bibfnamefont {T.}~\bibnamefont {Yu}},
  \ and\ \bibinfo {author} {\bibfnamefont {J.}~\bibnamefont {Eberly}},\
  }\href@noop {} {\bibfield  {journal} {\bibinfo  {journal} {J. Phys. B}\
  }\textbf {\bibinfo {volume} {39}},\ \bibinfo {pages} {S621} (\bibinfo {year}
  {2006})}\BibitemShut {NoStop}%
\bibitem [{\citenamefont {Berrada}\ \emph {et~al.}(2012)\citenamefont
  {Berrada}, \citenamefont {Fanchini},\ and\ \citenamefont
  {Abdel-Khalek}}]{berrada2012quantum}%
  \BibitemOpen
  \bibfield  {author} {\bibinfo {author} {\bibfnamefont {K.}~\bibnamefont
  {Berrada}}, \bibinfo {author} {\bibfnamefont {F.~F.}\ \bibnamefont
  {Fanchini}}, \ and\ \bibinfo {author} {\bibfnamefont {S.}~\bibnamefont
  {Abdel-Khalek}},\ }\href@noop {} {\bibfield  {journal} {\bibinfo  {journal}
  {Phys. Rev. A}\ }\textbf {\bibinfo {volume} {85}},\ \bibinfo {pages} {052315}
  (\bibinfo {year} {2012})}\BibitemShut {NoStop}%
\bibitem [{\citenamefont {Mohamed}\ \emph {et~al.}(2019)\citenamefont
  {Mohamed}, \citenamefont {Eleuch},\ and\ \citenamefont
  {Ooi}}]{mohamed2019non}%
  \BibitemOpen
  \bibfield  {author} {\bibinfo {author} {\bibfnamefont {A.-B.}\ \bibnamefont
  {Mohamed}}, \bibinfo {author} {\bibfnamefont {H.}~\bibnamefont {Eleuch}}, \
  and\ \bibinfo {author} {\bibfnamefont {C.~R.}\ \bibnamefont {Ooi}},\
  }\href@noop {} {\bibfield  {journal} {\bibinfo  {journal} {Sci. Rep.}\
  }\textbf {\bibinfo {volume} {9}},\ \bibinfo {pages} {1} (\bibinfo {year}
  {2019})}\BibitemShut {NoStop}%
\bibitem [{\citenamefont {Peres}(1996)}]{peres1996separability}%
  \BibitemOpen
  \bibfield  {author} {\bibinfo {author} {\bibfnamefont {A.}~\bibnamefont
  {Peres}},\ }\href@noop {} {\bibfield  {journal} {\bibinfo  {journal} {Phys.
  Rev. Lett.}\ }\textbf {\bibinfo {volume} {77}},\ \bibinfo {pages} {1413}
  (\bibinfo {year} {1996})}\BibitemShut {NoStop}%
\bibitem [{\citenamefont {Horodecki}\ \emph {et~al.}(1996)\citenamefont
  {Horodecki}, \citenamefont {Horodecki},\ and\ \citenamefont
  {Horodecki}}]{horodeckis1996separability}%
  \BibitemOpen
  \bibfield  {author} {\bibinfo {author} {\bibfnamefont {M.}~\bibnamefont
  {Horodecki}}, \bibinfo {author} {\bibfnamefont {P.}~\bibnamefont
  {Horodecki}}, \ and\ \bibinfo {author} {\bibfnamefont {R.}~\bibnamefont
  {Horodecki}},\ }\href@noop {} {\bibfield  {journal} {\bibinfo  {journal}
  {Phys. Lett. A}\ }\textbf {\bibinfo {volume} {223}},\ \bibinfo {pages} {1}
  (\bibinfo {year} {1996})}\BibitemShut {NoStop}%
\bibitem [{\citenamefont {Horodecki}(1997)}]{horodecki1997separability}%
  \BibitemOpen
  \bibfield  {author} {\bibinfo {author} {\bibfnamefont {P.}~\bibnamefont
  {Horodecki}},\ }\href@noop {} {\bibfield  {journal} {\bibinfo  {journal}
  {Phys. Lett. A}\ }\textbf {\bibinfo {volume} {232}},\ \bibinfo {pages} {333}
  (\bibinfo {year} {1997})}\BibitemShut {NoStop}%
\bibitem [{\citenamefont {Brandao}(2005)}]{brandao2005quantifying}%
  \BibitemOpen
  \bibfield  {author} {\bibinfo {author} {\bibfnamefont {F.~G.}\ \bibnamefont
  {Brandao}},\ }\href@noop {} {\bibfield  {journal} {\bibinfo  {journal} {Phys.
  Rev. A}\ }\textbf {\bibinfo {volume} {72}},\ \bibinfo {pages} {022310}
  (\bibinfo {year} {2005})}\BibitemShut {NoStop}%
\bibitem [{\citenamefont {G{\"u}hne}(2004)}]{guhne2004characterizing}%
  \BibitemOpen
  \bibfield  {author} {\bibinfo {author} {\bibfnamefont {O.}~\bibnamefont
  {G{\"u}hne}},\ }\href@noop {} {\bibfield  {journal} {\bibinfo  {journal}
  {Phys. Rev. Lett.}\ }\textbf {\bibinfo {volume} {92}},\ \bibinfo {pages}
  {117903} (\bibinfo {year} {2004})}\BibitemShut {NoStop}%
\bibitem [{\citenamefont {G{\"u}hne}\ \emph {et~al.}(2007)\citenamefont
  {G{\"u}hne}, \citenamefont {Hyllus}, \citenamefont {Gittsovich},\ and\
  \citenamefont {Eisert}}]{guhne2007covariance}%
  \BibitemOpen
  \bibfield  {author} {\bibinfo {author} {\bibfnamefont {O.}~\bibnamefont
  {G{\"u}hne}}, \bibinfo {author} {\bibfnamefont {P.}~\bibnamefont {Hyllus}},
  \bibinfo {author} {\bibfnamefont {O.}~\bibnamefont {Gittsovich}}, \ and\
  \bibinfo {author} {\bibfnamefont {J.}~\bibnamefont {Eisert}},\ }\href@noop {}
  {\bibfield  {journal} {\bibinfo  {journal} {Phys. Rev. Lett.}\ }\textbf
  {\bibinfo {volume} {99}},\ \bibinfo {pages} {130504} (\bibinfo {year}
  {2007})}\BibitemShut {NoStop}%
\bibitem [{\citenamefont {G{\"u}hne}\ \emph {et~al.}(2006)\citenamefont
  {G{\"u}hne}, \citenamefont {Mechler}, \citenamefont {T{\'o}th},\ and\
  \citenamefont {Adam}}]{guhne2006entanglement}%
  \BibitemOpen
  \bibfield  {author} {\bibinfo {author} {\bibfnamefont {O.}~\bibnamefont
  {G{\"u}hne}}, \bibinfo {author} {\bibfnamefont {M.}~\bibnamefont {Mechler}},
  \bibinfo {author} {\bibfnamefont {G.}~\bibnamefont {T{\'o}th}}, \ and\
  \bibinfo {author} {\bibfnamefont {P.}~\bibnamefont {Adam}},\ }\href@noop {}
  {\bibfield  {journal} {\bibinfo  {journal} {Phys. Rev. A}\ }\textbf {\bibinfo
  {volume} {74}},\ \bibinfo {pages} {010301} (\bibinfo {year}
  {2006})}\BibitemShut {NoStop}%
\bibitem [{\citenamefont {Gittsovich}\ \emph {et~al.}(2008)\citenamefont
  {Gittsovich}, \citenamefont {G{\"u}hne}, \citenamefont {Hyllus},\ and\
  \citenamefont {Eisert}}]{gittsovich2008unifying}%
  \BibitemOpen
  \bibfield  {author} {\bibinfo {author} {\bibfnamefont {O.}~\bibnamefont
  {Gittsovich}}, \bibinfo {author} {\bibfnamefont {O.}~\bibnamefont
  {G{\"u}hne}}, \bibinfo {author} {\bibfnamefont {P.}~\bibnamefont {Hyllus}}, \
  and\ \bibinfo {author} {\bibfnamefont {J.}~\bibnamefont {Eisert}},\
  }\href@noop {} {\bibfield  {journal} {\bibinfo  {journal} {Phys. Rev. A}\
  }\textbf {\bibinfo {volume} {78}},\ \bibinfo {pages} {052319} (\bibinfo
  {year} {2008})}\BibitemShut {NoStop}%
\bibitem [{\citenamefont {Li}\ and\ \citenamefont
  {Qiao}(2018)}]{li2018necessary}%
  \BibitemOpen
  \bibfield  {author} {\bibinfo {author} {\bibfnamefont {J.-L.}\ \bibnamefont
  {Li}}\ and\ \bibinfo {author} {\bibfnamefont {C.-F.}\ \bibnamefont {Qiao}},\
  }\href@noop {} {\bibfield  {journal} {\bibinfo  {journal} {Sci. Rep.}\
  }\textbf {\bibinfo {volume} {8}},\ \bibinfo {pages} {1442} (\bibinfo {year}
  {2018})}\BibitemShut {NoStop}%
\bibitem [{\citenamefont {de~Vicente}(2007)}]{vicente2007separability}%
  \BibitemOpen
  \bibfield  {author} {\bibinfo {author} {\bibfnamefont {J.~I.}\ \bibnamefont
  {de~Vicente}},\ }\href@noop {} {\bibfield  {journal} {\bibinfo  {journal}
  {Quantum Inf. Comput.}\ }\textbf {\bibinfo {volume} {7}},\ \bibinfo {pages}
  {624–638} (\bibinfo {year} {2007})}\BibitemShut {NoStop}%
\bibitem [{\citenamefont {Carmi}\ and\ \citenamefont
  {Cohen}(2018)}]{carmi2018significance}%
  \BibitemOpen
  \bibfield  {author} {\bibinfo {author} {\bibfnamefont {A.}~\bibnamefont
  {Carmi}}\ and\ \bibinfo {author} {\bibfnamefont {E.}~\bibnamefont {Cohen}},\
  }\href@noop {} {\bibfield  {journal} {\bibinfo  {journal} {Entropy}\ }\textbf
  {\bibinfo {volume} {20}},\ \bibinfo {pages} {500} (\bibinfo {year}
  {2018})}\BibitemShut {NoStop}%
\bibitem [{\citenamefont {Carmi}\ and\ \citenamefont
  {Cohen}(2019)}]{carmi2019relativistic}%
  \BibitemOpen
  \bibfield  {author} {\bibinfo {author} {\bibfnamefont {A.}~\bibnamefont
  {Carmi}}\ and\ \bibinfo {author} {\bibfnamefont {E.}~\bibnamefont {Cohen}},\
  }\href@noop {} {\bibfield  {journal} {\bibinfo  {journal} {Sci. Adv.}\
  }\textbf {\bibinfo {volume} {5}},\ \bibinfo {pages} {eaav8370} (\bibinfo
  {year} {2019})}\BibitemShut {NoStop}%
\bibitem [{\citenamefont {Te'eni}\ \emph {et~al.}(2019)\citenamefont {Te'eni},
  \citenamefont {Peled}, \citenamefont {Cohen},\ and\ \citenamefont
  {Carmi}}]{te2019multiplicative}%
  \BibitemOpen
  \bibfield  {author} {\bibinfo {author} {\bibfnamefont {A.}~\bibnamefont
  {Te'eni}}, \bibinfo {author} {\bibfnamefont {B.~Y.}\ \bibnamefont {Peled}},
  \bibinfo {author} {\bibfnamefont {E.}~\bibnamefont {Cohen}}, \ and\ \bibinfo
  {author} {\bibfnamefont {A.}~\bibnamefont {Carmi}},\ }\href@noop {}
  {\bibfield  {journal} {\bibinfo  {journal} {Phys. Rev. A}\ }\textbf {\bibinfo
  {volume} {99}},\ \bibinfo {pages} {040102} (\bibinfo {year}
  {2019})}\BibitemShut {NoStop}%
\bibitem [{\citenamefont {Pozsgay}\ \emph {et~al.}(2017)\citenamefont
  {Pozsgay}, \citenamefont {Hirsch}, \citenamefont {Branciard},\ and\
  \citenamefont {Brunner}}]{Pozsgay}%
  \BibitemOpen
  \bibfield  {author} {\bibinfo {author} {\bibfnamefont {V.}~\bibnamefont
  {Pozsgay}}, \bibinfo {author} {\bibfnamefont {F.}~\bibnamefont {Hirsch}},
  \bibinfo {author} {\bibfnamefont {C.}~\bibnamefont {Branciard}}, \ and\
  \bibinfo {author} {\bibfnamefont {N.}~\bibnamefont {Brunner}},\ }\href@noop
  {} {\bibfield  {journal} {\bibinfo  {journal} {Phys. Rev. A}\ }\textbf
  {\bibinfo {volume} {96}},\ \bibinfo {pages} {062128} (\bibinfo {year}
  {2017})}\BibitemShut {NoStop}%
\bibitem [{\citenamefont {Ollivier}\ and\ \citenamefont
  {Zurek}(2001)}]{ollivier2001quantum}%
  \BibitemOpen
  \bibfield  {author} {\bibinfo {author} {\bibfnamefont {H.}~\bibnamefont
  {Ollivier}}\ and\ \bibinfo {author} {\bibfnamefont {W.~H.}\ \bibnamefont
  {Zurek}},\ }\href@noop {} {\bibfield  {journal} {\bibinfo  {journal} {Phys.
  Rev. Lett.}\ }\textbf {\bibinfo {volume} {88}},\ \bibinfo {pages} {017901}
  (\bibinfo {year} {2001})}\BibitemShut {NoStop}%
\bibitem [{\citenamefont {Zurek}(2000)}]{zurek2000einselection}%
  \BibitemOpen
  \bibfield  {author} {\bibinfo {author} {\bibfnamefont {W.~H.}\ \bibnamefont
  {Zurek}},\ }\href@noop {} {\bibfield  {journal} {\bibinfo  {journal} {Ann.
  Phys.}\ }\textbf {\bibinfo {volume} {9}},\ \bibinfo {pages} {855} (\bibinfo
  {year} {2000})}\BibitemShut {NoStop}%
\bibitem [{\citenamefont {Henderson}\ and\ \citenamefont
  {Vedral}(2001)}]{henderson2001classical}%
  \BibitemOpen
  \bibfield  {author} {\bibinfo {author} {\bibfnamefont {L.}~\bibnamefont
  {Henderson}}\ and\ \bibinfo {author} {\bibfnamefont {V.}~\bibnamefont
  {Vedral}},\ }\href@noop {} {\bibfield  {journal} {\bibinfo  {journal} {J.
  Phys. A}\ }\textbf {\bibinfo {volume} {34}},\ \bibinfo {pages} {6899}
  (\bibinfo {year} {2001})}\BibitemShut {NoStop}%
\bibitem [{\citenamefont {Giorda}\ and\ \citenamefont
  {Paris}(2010)}]{giorda2010gaussian}%
  \BibitemOpen
  \bibfield  {author} {\bibinfo {author} {\bibfnamefont {P.}~\bibnamefont
  {Giorda}}\ and\ \bibinfo {author} {\bibfnamefont {M.~G.}\ \bibnamefont
  {Paris}},\ }\href@noop {} {\bibfield  {journal} {\bibinfo  {journal} {Phys.
  Rev. Lett.}\ }\textbf {\bibinfo {volume} {105}},\ \bibinfo {pages} {020503}
  (\bibinfo {year} {2010})}\BibitemShut {NoStop}%
\bibitem [{\citenamefont {Luo}(2008)}]{luo2008quantum}%
  \BibitemOpen
  \bibfield  {author} {\bibinfo {author} {\bibfnamefont {S.}~\bibnamefont
  {Luo}},\ }\href@noop {} {\bibfield  {journal} {\bibinfo  {journal} {Phys.
  Rev. A}\ }\textbf {\bibinfo {volume} {77}},\ \bibinfo {pages} {042303}
  (\bibinfo {year} {2008})}\BibitemShut {NoStop}%
\bibitem [{\citenamefont {Bera}\ \emph {et~al.}(2017)\citenamefont {Bera},
  \citenamefont {Das}, \citenamefont {Sadhukhan}, \citenamefont {Roy},
  \citenamefont {De},\ and\ \citenamefont {Sen}}]{bera2017quantum}%
  \BibitemOpen
  \bibfield  {author} {\bibinfo {author} {\bibfnamefont {A.}~\bibnamefont
  {Bera}}, \bibinfo {author} {\bibfnamefont {T.}~\bibnamefont {Das}}, \bibinfo
  {author} {\bibfnamefont {D.}~\bibnamefont {Sadhukhan}}, \bibinfo {author}
  {\bibfnamefont {S.~S.}\ \bibnamefont {Roy}}, \bibinfo {author} {\bibfnamefont
  {A.~S.}\ \bibnamefont {De}}, \ and\ \bibinfo {author} {\bibfnamefont
  {U.}~\bibnamefont {Sen}},\ }\href@noop {} {\bibfield  {journal} {\bibinfo
  {journal} {Rep. Prog. Phys.}\ }\textbf {\bibinfo {volume} {81}},\ \bibinfo
  {pages} {024001} (\bibinfo {year} {2017})}\BibitemShut {NoStop}%
\bibitem [{\citenamefont {Daki{\'c}}\ \emph {et~al.}(2010)\citenamefont
  {Daki{\'c}}, \citenamefont {Vedral},\ and\ \citenamefont
  {Brukner}}]{dakic2010necessary}%
  \BibitemOpen
  \bibfield  {author} {\bibinfo {author} {\bibfnamefont {B.}~\bibnamefont
  {Daki{\'c}}}, \bibinfo {author} {\bibfnamefont {V.}~\bibnamefont {Vedral}}, \
  and\ \bibinfo {author} {\bibfnamefont {{\v{C}}.}~\bibnamefont {Brukner}},\
  }\href@noop {} {\bibfield  {journal} {\bibinfo  {journal} {Phys. Rev. Lett.}\
  }\textbf {\bibinfo {volume} {105}},\ \bibinfo {pages} {190502} (\bibinfo
  {year} {2010})}\BibitemShut {NoStop}%
\bibitem [{\citenamefont {Luo}\ and\ \citenamefont
  {Fu}(2010)}]{luo2010geometric}%
  \BibitemOpen
  \bibfield  {author} {\bibinfo {author} {\bibfnamefont {S.}~\bibnamefont
  {Luo}}\ and\ \bibinfo {author} {\bibfnamefont {S.}~\bibnamefont {Fu}},\
  }\href@noop {} {\bibfield  {journal} {\bibinfo  {journal} {Phys. Rev. A}\
  }\textbf {\bibinfo {volume} {82}},\ \bibinfo {pages} {034302} (\bibinfo
  {year} {2010})}\BibitemShut {NoStop}%
\bibitem [{\citenamefont {Lupo}\ \emph {et~al.}(2008)\citenamefont {Lupo},
  \citenamefont {Aniello},\ and\ \citenamefont
  {Scardicchio}}]{lupo2008bipartite}%
  \BibitemOpen
  \bibfield  {author} {\bibinfo {author} {\bibfnamefont {C.}~\bibnamefont
  {Lupo}}, \bibinfo {author} {\bibfnamefont {P.}~\bibnamefont {Aniello}}, \
  and\ \bibinfo {author} {\bibfnamefont {A.}~\bibnamefont {Scardicchio}},\
  }\href@noop {} {\bibfield  {journal} {\bibinfo  {journal} {J. Phys. A}\
  }\textbf {\bibinfo {volume} {41}},\ \bibinfo {pages} {415301} (\bibinfo
  {year} {2008})}\BibitemShut {NoStop}%
\bibitem [{\citenamefont {Li}\ \emph {et~al.}(2011)\citenamefont {Li},
  \citenamefont {Poon},\ and\ \citenamefont {Sze}}]{li2011note}%
  \BibitemOpen
  \bibfield  {author} {\bibinfo {author} {\bibfnamefont {C.-K.}\ \bibnamefont
  {Li}}, \bibinfo {author} {\bibfnamefont {Y.-T.}\ \bibnamefont {Poon}}, \ and\
  \bibinfo {author} {\bibfnamefont {N.-S.}\ \bibnamefont {Sze}},\ }\href@noop
  {} {\bibfield  {journal} {\bibinfo  {journal} {J. Phys. A}\ }\textbf
  {\bibinfo {volume} {44}},\ \bibinfo {pages} {315304} (\bibinfo {year}
  {2011})}\BibitemShut {NoStop}%
\bibitem [{\citenamefont {Chen}\ and\ \citenamefont
  {Wu}(2002)}]{chen2002matrix}%
  \BibitemOpen
  \bibfield  {author} {\bibinfo {author} {\bibfnamefont {K.}~\bibnamefont
  {Chen}}\ and\ \bibinfo {author} {\bibfnamefont {L.-A.}\ \bibnamefont {Wu}},\
  }\href@noop {} {\bibfield  {journal} {\bibinfo  {journal} {Quantum Inf.
  Comput.}\ }\textbf {\bibinfo {volume} {3}} (\bibinfo {year}
  {2002})}\BibitemShut {NoStop}%
\bibitem [{\citenamefont {Rudolph}(2005)}]{rudolph2005further}%
  \BibitemOpen
  \bibfield  {author} {\bibinfo {author} {\bibfnamefont {O.}~\bibnamefont
  {Rudolph}},\ }\href@noop {} {\bibfield  {journal} {\bibinfo  {journal}
  {Quantum Inf. Process.}\ }\textbf {\bibinfo {volume} {4}},\ \bibinfo {pages}
  {219} (\bibinfo {year} {2005})}\BibitemShut {NoStop}%
\bibitem [{\citenamefont {Simon}(2000)}]{simon2000peres}%
  \BibitemOpen
  \bibfield  {author} {\bibinfo {author} {\bibfnamefont {R.}~\bibnamefont
  {Simon}},\ }\href@noop {} {\bibfield  {journal} {\bibinfo  {journal} {Phys.
  Rev. Lett.}\ }\textbf {\bibinfo {volume} {84}},\ \bibinfo {pages} {2726}
  (\bibinfo {year} {2000})}\BibitemShut {NoStop}%
\bibitem [{\citenamefont {Dodonov}\ \emph {et~al.}(2004)\citenamefont
  {Dodonov}, \citenamefont {Dodonov},\ and\ \citenamefont
  {Mizrahi}}]{dodonov2004separability}%
  \BibitemOpen
  \bibfield  {author} {\bibinfo {author} {\bibfnamefont {A.}~\bibnamefont
  {Dodonov}}, \bibinfo {author} {\bibfnamefont {V.}~\bibnamefont {Dodonov}}, \
  and\ \bibinfo {author} {\bibfnamefont {S.}~\bibnamefont {Mizrahi}},\
  }\href@noop {} {\bibfield  {journal} {\bibinfo  {journal} {J. Phys. A}\
  }\textbf {\bibinfo {volume} {38}},\ \bibinfo {pages} {683} (\bibinfo {year}
  {2004})}\BibitemShut {NoStop}%
\bibitem [{\citenamefont {De~Castro}\ and\ \citenamefont
  {Dodonov}(2006)}]{de2006purity}%
  \BibitemOpen
  \bibfield  {author} {\bibinfo {author} {\bibfnamefont {A.}~\bibnamefont
  {De~Castro}}\ and\ \bibinfo {author} {\bibfnamefont {V.}~\bibnamefont
  {Dodonov}},\ }\href@noop {} {\bibfield  {journal} {\bibinfo  {journal} {Phys.
  Rev. A}\ }\textbf {\bibinfo {volume} {73}},\ \bibinfo {pages} {065801}
  (\bibinfo {year} {2006})}\BibitemShut {NoStop}%
\bibitem [{\citenamefont {G{\"u}hne}\ and\ \citenamefont
  {T{\'o}th}(2009)}]{guhne2009entanglement}%
  \BibitemOpen
  \bibfield  {author} {\bibinfo {author} {\bibfnamefont {O.}~\bibnamefont
  {G{\"u}hne}}\ and\ \bibinfo {author} {\bibfnamefont {G.}~\bibnamefont
  {T{\'o}th}},\ }\href@noop {} {\bibfield  {journal} {\bibinfo  {journal}
  {Phys. Rep.}\ }\textbf {\bibinfo {volume} {474}},\ \bibinfo {pages} {1}
  (\bibinfo {year} {2009})}\BibitemShut {NoStop}%
\bibitem [{\citenamefont {Kent}\ \emph {et~al.}(1999)\citenamefont {Kent},
  \citenamefont {Linden},\ and\ \citenamefont {Massar}}]{kent1999optimal}%
  \BibitemOpen
  \bibfield  {author} {\bibinfo {author} {\bibfnamefont {A.}~\bibnamefont
  {Kent}}, \bibinfo {author} {\bibfnamefont {N.}~\bibnamefont {Linden}}, \ and\
  \bibinfo {author} {\bibfnamefont {S.}~\bibnamefont {Massar}},\ }\href@noop {}
  {\bibfield  {journal} {\bibinfo  {journal} {Phys. Rev. Lett.}\ }\textbf
  {\bibinfo {volume} {83}},\ \bibinfo {pages} {2656} (\bibinfo {year}
  {1999})}\BibitemShut {NoStop}%
\bibitem [{\citenamefont {Verstraete}\ \emph {et~al.}(2003)\citenamefont
  {Verstraete}, \citenamefont {Dehaene},\ and\ \citenamefont
  {De~Moor}}]{verstraete2003normal}%
  \BibitemOpen
  \bibfield  {author} {\bibinfo {author} {\bibfnamefont {F.}~\bibnamefont
  {Verstraete}}, \bibinfo {author} {\bibfnamefont {J.}~\bibnamefont {Dehaene}},
  \ and\ \bibinfo {author} {\bibfnamefont {B.}~\bibnamefont {De~Moor}},\
  }\href@noop {} {\bibfield  {journal} {\bibinfo  {journal} {Phys. Rev. A}\
  }\textbf {\bibinfo {volume} {68}},\ \bibinfo {pages} {012103} (\bibinfo
  {year} {2003})}\BibitemShut {NoStop}%
\bibitem [{\citenamefont {Leinaas}\ \emph {et~al.}(2006)\citenamefont
  {Leinaas}, \citenamefont {Myrheim},\ and\ \citenamefont
  {Ovrum}}]{leinaas2006geometrical}%
  \BibitemOpen
  \bibfield  {author} {\bibinfo {author} {\bibfnamefont {J.~M.}\ \bibnamefont
  {Leinaas}}, \bibinfo {author} {\bibfnamefont {J.}~\bibnamefont {Myrheim}}, \
  and\ \bibinfo {author} {\bibfnamefont {E.}~\bibnamefont {Ovrum}},\
  }\href@noop {} {\bibfield  {journal} {\bibinfo  {journal} {Phys. Rev. A}\
  }\textbf {\bibinfo {volume} {74}},\ \bibinfo {pages} {012313} (\bibinfo
  {year} {2006})}\BibitemShut {NoStop}%
\bibitem [{\citenamefont {Zauner}(2011)}]{zauner2011quantum}%
  \BibitemOpen
  \bibfield  {author} {\bibinfo {author} {\bibfnamefont {G.}~\bibnamefont
  {Zauner}},\ }\href@noop {} {\bibfield  {journal} {\bibinfo  {journal} {Int.
  J. Quantum Inf.}\ }\textbf {\bibinfo {volume} {9}},\ \bibinfo {pages} {445}
  (\bibinfo {year} {2011})}\BibitemShut {NoStop}%
\bibitem [{\citenamefont {Renes}\ \emph {et~al.}(2004)\citenamefont {Renes},
  \citenamefont {Blume-Kohout}, \citenamefont {Scott},\ and\ \citenamefont
  {Caves}}]{renes2004symmetric}%
  \BibitemOpen
  \bibfield  {author} {\bibinfo {author} {\bibfnamefont {J.~M.}\ \bibnamefont
  {Renes}}, \bibinfo {author} {\bibfnamefont {R.}~\bibnamefont {Blume-Kohout}},
  \bibinfo {author} {\bibfnamefont {A.~J.}\ \bibnamefont {Scott}}, \ and\
  \bibinfo {author} {\bibfnamefont {C.~M.}\ \bibnamefont {Caves}},\ }\href@noop
  {} {\bibfield  {journal} {\bibinfo  {journal} {J. Math. Phys.}\ }\textbf
  {\bibinfo {volume} {45}},\ \bibinfo {pages} {2171} (\bibinfo {year}
  {2004})}\BibitemShut {NoStop}%
\bibitem [{\citenamefont {Virz{\`\i}}\ \emph {et~al.}(2019)\citenamefont
  {Virz{\`\i}}, \citenamefont {Rebufello}, \citenamefont {Avella},
  \citenamefont {Piacentini}, \citenamefont {Gramegna}, \citenamefont
  {Berchera}, \citenamefont {Degiovanni},\ and\ \citenamefont
  {Genovese}}]{virzi2019optimal}%
  \BibitemOpen
  \bibfield  {author} {\bibinfo {author} {\bibfnamefont {S.}~\bibnamefont
  {Virz{\`\i}}}, \bibinfo {author} {\bibfnamefont {E.}~\bibnamefont
  {Rebufello}}, \bibinfo {author} {\bibfnamefont {A.}~\bibnamefont {Avella}},
  \bibinfo {author} {\bibfnamefont {F.}~\bibnamefont {Piacentini}}, \bibinfo
  {author} {\bibfnamefont {M.}~\bibnamefont {Gramegna}}, \bibinfo {author}
  {\bibfnamefont {I.~R.}\ \bibnamefont {Berchera}}, \bibinfo {author}
  {\bibfnamefont {I.~P.}\ \bibnamefont {Degiovanni}}, \ and\ \bibinfo {author}
  {\bibfnamefont {M.}~\bibnamefont {Genovese}},\ }\href@noop {} {\bibfield
  {journal} {\bibinfo  {journal} {Sci. Rep.}\ }\textbf {\bibinfo {volume}
  {9}},\ \bibinfo {pages} {1} (\bibinfo {year} {2019})}\BibitemShut {NoStop}%
\bibitem [{\citenamefont {Galve}\ \emph {et~al.}(2011)\citenamefont {Galve},
  \citenamefont {Giorgi},\ and\ \citenamefont {Zambrini}}]{galve2011maximally}%
  \BibitemOpen
  \bibfield  {author} {\bibinfo {author} {\bibfnamefont {F.}~\bibnamefont
  {Galve}}, \bibinfo {author} {\bibfnamefont {G.~L.}\ \bibnamefont {Giorgi}}, \
  and\ \bibinfo {author} {\bibfnamefont {R.}~\bibnamefont {Zambrini}},\
  }\href@noop {} {\bibfield  {journal} {\bibinfo  {journal} {Phys. Rev. A}\
  }\textbf {\bibinfo {volume} {83}},\ \bibinfo {pages} {012102} (\bibinfo
  {year} {2011})}\BibitemShut {NoStop}%
\bibitem [{\citenamefont {Ye}\ \emph {et~al.}(2013)\citenamefont {Ye},
  \citenamefont {Liu}, \citenamefont {Chen}, \citenamefont {Liu},\ and\
  \citenamefont {Zhang}}]{ye2013analytic}%
  \BibitemOpen
  \bibfield  {author} {\bibinfo {author} {\bibfnamefont {B.}~\bibnamefont
  {Ye}}, \bibinfo {author} {\bibfnamefont {Y.}~\bibnamefont {Liu}}, \bibinfo
  {author} {\bibfnamefont {J.}~\bibnamefont {Chen}}, \bibinfo {author}
  {\bibfnamefont {X.}~\bibnamefont {Liu}}, \ and\ \bibinfo {author}
  {\bibfnamefont {Z.}~\bibnamefont {Zhang}},\ }\href@noop {} {\bibfield
  {journal} {\bibinfo  {journal} {Quantum Inf. Process.}\ }\textbf {\bibinfo
  {volume} {12}},\ \bibinfo {pages} {2355} (\bibinfo {year}
  {2013})}\BibitemShut {NoStop}%
\bibitem [{\citenamefont {Piani}(2012)}]{piani2012problem}%
  \BibitemOpen
  \bibfield  {author} {\bibinfo {author} {\bibfnamefont {M.}~\bibnamefont
  {Piani}},\ }\href@noop {} {\bibfield  {journal} {\bibinfo  {journal} {Phys.
  Rev. A}\ }\textbf {\bibinfo {volume} {86}},\ \bibinfo {pages} {034101}
  (\bibinfo {year} {2012})}\BibitemShut {NoStop}%
\bibitem [{\citenamefont {Tufarelli}\ \emph {et~al.}(2012)\citenamefont
  {Tufarelli}, \citenamefont {Girolami}, \citenamefont {Vasile}, \citenamefont
  {Bose},\ and\ \citenamefont {Adesso}}]{tufarelli2012quantum}%
  \BibitemOpen
  \bibfield  {author} {\bibinfo {author} {\bibfnamefont {T.}~\bibnamefont
  {Tufarelli}}, \bibinfo {author} {\bibfnamefont {D.}~\bibnamefont {Girolami}},
  \bibinfo {author} {\bibfnamefont {R.}~\bibnamefont {Vasile}}, \bibinfo
  {author} {\bibfnamefont {S.}~\bibnamefont {Bose}}, \ and\ \bibinfo {author}
  {\bibfnamefont {G.}~\bibnamefont {Adesso}},\ }\href@noop {} {\bibfield
  {journal} {\bibinfo  {journal} {Phys. Rev. A}\ }\textbf {\bibinfo {volume}
  {86}},\ \bibinfo {pages} {052326} (\bibinfo {year} {2012})}\BibitemShut
  {NoStop}%
\bibitem [{\citenamefont {Paula}\ \emph {et~al.}(2013)\citenamefont {Paula},
  \citenamefont {de~Oliveira},\ and\ \citenamefont
  {Sarandy}}]{paula2013geometric}%
  \BibitemOpen
  \bibfield  {author} {\bibinfo {author} {\bibfnamefont {F.}~\bibnamefont
  {Paula}}, \bibinfo {author} {\bibfnamefont {T.~R.}\ \bibnamefont
  {de~Oliveira}}, \ and\ \bibinfo {author} {\bibfnamefont {M.}~\bibnamefont
  {Sarandy}},\ }\href@noop {} {\bibfield  {journal} {\bibinfo  {journal} {Phys.
  Rev. A}\ }\textbf {\bibinfo {volume} {87}},\ \bibinfo {pages} {064101}
  (\bibinfo {year} {2013})}\BibitemShut {NoStop}%
\bibitem [{\citenamefont {Roga}\ \emph {et~al.}(2016)\citenamefont {Roga},
  \citenamefont {Spehner},\ and\ \citenamefont
  {Illuminati}}]{roga2016geometric}%
  \BibitemOpen
  \bibfield  {author} {\bibinfo {author} {\bibfnamefont {W.}~\bibnamefont
  {Roga}}, \bibinfo {author} {\bibfnamefont {D.}~\bibnamefont {Spehner}}, \
  and\ \bibinfo {author} {\bibfnamefont {F.}~\bibnamefont {Illuminati}},\
  }\href@noop {} {\bibfield  {journal} {\bibinfo  {journal} {J. Phys. A}\
  }\textbf {\bibinfo {volume} {49}},\ \bibinfo {pages} {235301} (\bibinfo
  {year} {2016})}\BibitemShut {NoStop}%
\bibitem [{\citenamefont {de~Vicente}\ and\ \citenamefont
  {Huber}(2011)}]{de2011multipartite}%
  \BibitemOpen
  \bibfield  {author} {\bibinfo {author} {\bibfnamefont {J.~I.}\ \bibnamefont
  {de~Vicente}}\ and\ \bibinfo {author} {\bibfnamefont {M.}~\bibnamefont
  {Huber}},\ }\href@noop {} {\bibfield  {journal} {\bibinfo  {journal} {Phys.
  Rev. A}\ }\textbf {\bibinfo {volume} {84}},\ \bibinfo {pages} {062306}
  (\bibinfo {year} {2011})}\BibitemShut {NoStop}%
\bibitem [{\citenamefont {Jaeger}\ \emph {et~al.}(1993)\citenamefont {Jaeger},
  \citenamefont {Horne},\ and\ \citenamefont
  {Shimony}}]{jaeger1993complementarity}%
  \BibitemOpen
  \bibfield  {author} {\bibinfo {author} {\bibfnamefont {G.}~\bibnamefont
  {Jaeger}}, \bibinfo {author} {\bibfnamefont {M.~A.}\ \bibnamefont {Horne}}, \
  and\ \bibinfo {author} {\bibfnamefont {A.}~\bibnamefont {Shimony}},\ }\href
  {\doibase 10.1103/PhysRevA.48.1023} {\bibfield  {journal} {\bibinfo
  {journal} {Phys. Rev. A}\ }\textbf {\bibinfo {volume} {48}},\ \bibinfo
  {pages} {1023} (\bibinfo {year} {1993})}\BibitemShut {NoStop}%
\bibitem [{\citenamefont {Greenberger}\ and\ \citenamefont
  {Yasin}(1988)}]{greenberger1988simultaneous}%
  \BibitemOpen
  \bibfield  {author} {\bibinfo {author} {\bibfnamefont {D.~M.}\ \bibnamefont
  {Greenberger}}\ and\ \bibinfo {author} {\bibfnamefont {A.}~\bibnamefont
  {Yasin}},\ }\href {\doibase 10.1016/0375-9601(88)90114-4} {\bibfield
  {journal} {\bibinfo  {journal} {Phys. Lett. A}\ }\textbf {\bibinfo {volume}
  {128}},\ \bibinfo {pages} {391} (\bibinfo {year} {1988})}\BibitemShut
  {NoStop}%
\bibitem [{\citenamefont {Jaeger}\ \emph {et~al.}(1995)\citenamefont {Jaeger},
  \citenamefont {Shimony},\ and\ \citenamefont {Vaidman}}]{jaeger1995two}%
  \BibitemOpen
  \bibfield  {author} {\bibinfo {author} {\bibfnamefont {G.}~\bibnamefont
  {Jaeger}}, \bibinfo {author} {\bibfnamefont {A.}~\bibnamefont {Shimony}}, \
  and\ \bibinfo {author} {\bibfnamefont {L.}~\bibnamefont {Vaidman}},\ }\href
  {\doibase 10.1103/PhysRevA.51.54} {\bibfield  {journal} {\bibinfo  {journal}
  {Phys. Rev. A}\ }\textbf {\bibinfo {volume} {51}},\ \bibinfo {pages} {54}
  (\bibinfo {year} {1995})}\BibitemShut {NoStop}%
\bibitem [{\citenamefont {Englert}(1996)}]{englert1996fringe}%
  \BibitemOpen
  \bibfield  {author} {\bibinfo {author} {\bibfnamefont {B.-G.}\ \bibnamefont
  {Englert}},\ }\href {\doibase 10.1103/PhysRevLett.77.2154} {\bibfield
  {journal} {\bibinfo  {journal} {Phys. Rev. Lett.}\ }\textbf {\bibinfo
  {volume} {77}},\ \bibinfo {pages} {2154} (\bibinfo {year}
  {1996})}\BibitemShut {NoStop}%
\bibitem [{\citenamefont {Franson}(1989)}]{franson1989bell}%
  \BibitemOpen
  \bibfield  {author} {\bibinfo {author} {\bibfnamefont {J.~D.}\ \bibnamefont
  {Franson}},\ }\href {\doibase 10.1103/PhysRevLett.62.2205} {\bibfield
  {journal} {\bibinfo  {journal} {Phys. Rev. Lett.}\ }\textbf {\bibinfo
  {volume} {62}},\ \bibinfo {pages} {2205} (\bibinfo {year}
  {1989})}\BibitemShut {NoStop}%
\bibitem [{\citenamefont {Peled}\ \emph {et~al.}(2020)\citenamefont {Peled},
  \citenamefont {Te’eni}, \citenamefont {Georgiev}, \citenamefont {Cohen},\
  and\ \citenamefont {Carmi}}]{peled2020double}%
  \BibitemOpen
  \bibfield  {author} {\bibinfo {author} {\bibfnamefont {B.~Y.}\ \bibnamefont
  {Peled}}, \bibinfo {author} {\bibfnamefont {A.}~\bibnamefont {Te’eni}},
  \bibinfo {author} {\bibfnamefont {D.}~\bibnamefont {Georgiev}}, \bibinfo
  {author} {\bibfnamefont {E.}~\bibnamefont {Cohen}}, \ and\ \bibinfo {author}
  {\bibfnamefont {A.}~\bibnamefont {Carmi}},\ }\href@noop {} {\bibfield
  {journal} {\bibinfo  {journal} {Appl. Sci.}\ }\textbf {\bibinfo {volume}
  {10}},\ \bibinfo {pages} {792} (\bibinfo {year} {2020})}\BibitemShut
  {NoStop}%
\bibitem [{\citenamefont {Horn}\ and\ \citenamefont
  {Johnson}(2012)}]{horn2012matrix}%
  \BibitemOpen
  \bibfield  {author} {\bibinfo {author} {\bibfnamefont {R.~A.}\ \bibnamefont
  {Horn}}\ and\ \bibinfo {author} {\bibfnamefont {C.~R.}\ \bibnamefont
  {Johnson}},\ }\href@noop {} {\emph {\bibinfo {title} {Matrix analysis}}}\
  (\bibinfo  {publisher} {Cambridge university press},\ \bibinfo {year}
  {2012})\BibitemShut {NoStop}%
\bibitem [{\citenamefont {Waldron}(2018)}]{waldron2018introduction}%
  \BibitemOpen
  \bibfield  {author} {\bibinfo {author} {\bibfnamefont {S.~F.}\ \bibnamefont
  {Waldron}},\ }\href@noop {} {\emph {\bibinfo {title} {An introduction to
  finite tight frames}}}\ (\bibinfo  {publisher} {Springer},\ \bibinfo {year}
  {2018})\BibitemShut {NoStop}%
\bibitem [{\citenamefont {Hiai}\ and\ \citenamefont
  {Petz}(2014)}]{hiai2014introduction}%
  \BibitemOpen
  \bibfield  {author} {\bibinfo {author} {\bibfnamefont {F.}~\bibnamefont
  {Hiai}}\ and\ \bibinfo {author} {\bibfnamefont {D.}~\bibnamefont {Petz}},\
  }\href@noop {} {\emph {\bibinfo {title} {Introduction to matrix analysis and
  applications}}}\ (\bibinfo  {publisher} {Springer Science \& Business
  Media},\ \bibinfo {year} {2014})\BibitemShut {NoStop}%
\end{thebibliography}%

\end{document}